\documentclass[twocolumn]{aastex62}

\usepackage{amsmath}
\usepackage{csquotes}
\usepackage{bm}
\usepackage{savesym}

\savesymbol{tablenum} 
\usepackage{siunitx}
\restoresymbol{SI}{tablenum}

\usepackage{grfext}
\graphicspath{{./}{figures/}}  

\newcommand{\Ha}{H$\alpha$~}
\newcommand{\hii}{H{\sc ii}~}
\newcommand{\hi}{H{\sc i}~}

\DeclareSIUnit\arcsec{arcsec}
\DeclareSIUnit\arcmin{arcmin}
\DeclareSIUnit\cMpc{cMpc}  
\DeclareSIUnit\cGpc{cGpc}  
\DeclareSIUnit\deg{deg}
\DeclareSIUnit\erg{erg}
\DeclareSIUnit\gauss{G}
\DeclareSIUnit\hour{hr}  
\DeclareSIUnit\hubble{\text{$h$}}
\DeclareSIUnit\jansky{Jy}
\DeclareSIUnit\parsec{pc}
\DeclareSIUnit\rayleigh{Rayleigh}
\DeclareSIUnit\solarmass{\text{M$_{\odot}$}}
\DeclareSIUnit\Msun{\solarmass}

\DeclareSIUnit\keV{\kilo\electronvolt}
\DeclareSIUnit\kpc{\kilo\parsec}
\DeclareSIUnit\pc{\parsec}
\DeclareSIUnit\mJy{\milli\jansky}
\DeclareSIUnit\mK{\milli\kelvin}
\DeclareSIUnit\uG{\micro\gauss}
\DeclareSIUnit\Gyr{\giga\year}
\DeclareSIUnit\MHz{\mega\hertz}
\DeclareSIUnit\Mpc{\mega\parsec}
\DeclareSIUnit\Gpc{\giga\parsec}

\def\equationautorefname~#1\null{Equation~(#1)\null}

\newcounter{sssseccount}

\shorttitle{Modeling the Galactic Free-free Emission}
\shortauthors{Lian~\textit{et~al.}}

\begin{document}

\title{%
  A Monte Carlo Implementation of Galactic Free--Free Emission for the EoR Foreground Models
}

\author[0000-0002-8516-2150]{Xiaoli Lian}
\email{lianxiaoli87@sjtu.edu.cn}
\affiliation{School of Physics and Astronomy,
  Shanghai Jiao Tong University,
  800 Dongchuan Road, Shanghai 200240, People’s Republic of China}

\author{Haiguang Xu}
\email{hgxu@sjtu.edu.cn}
\affiliation{School of Physics and Astronomy,
  Shanghai Jiao Tong University,
  800 Dongchuan Road, Shanghai 200240, People’s Republic of China}
\affiliation{Tsung-Dao Lee Institute,
  Shanghai Jiao Tong University,
  800 Dongchuan Road, Shanghai 200240, People’s Republic of China}
\affiliation{IFSA Collaborative Innovation Center,
  Shanghai Jiao Tong University,
  800 Dongchuan Road, Shanghai 200240, People’s Republic of China}

\author{Dongchao Zheng}
\affiliation{School of Physics and Astronomy,
	Shanghai Jiao Tong University,
	800 Dongchuan Road, Shanghai 200240, People’s Republic of China}
\author{Zhenghao Zhu}
\affiliation{School of Physics and Astronomy,
	Shanghai Jiao Tong University,
	800 Dongchuan Road, Shanghai 200240, People’s Republic of China}
\author{Dan Hu}
\affiliation{School of Physics and Astronomy,
	Shanghai Jiao Tong University,
	800 Dongchuan Road, Shanghai 200240, People’s Republic of China}



\begin{abstract}
\noindent

The overwhelming foreground causes severe contamination on the detection of 21 cm signal during the Epoch of Reionization (EoR). Among various foreground components, the Galactic free--free emission is less studied, so that its impact on the EoR observation remains unclear. To better constrain this emission, we perform Monte Carlo simulation of \Ha emission, which comprises direct and scattered \Ha radiation from \hii regions and warm ionized medium (WIM). The positions and radii of \hii regions are quoted from the Wide-Field Infrared Survey Explorer \hii catalog, and the WIM is described by an axisymmetric model. The scattering is off dust and free electrons that are realized by applying an exponential fitting to the HI4PI \hi map and an exponential disk model, respectively. The simulated \Ha intensity, the \texttt{Simfast21} software, and the latest SKA1-Low layout configuration are employed to simulate the SKA \enquote{observed} images of Galactic free--free emission and the EoR signal. By analyzing the one-dimensional power spectra, we find that the Galactic free--free emission can be about \numrange{e5.4}{e2.1}, \numrange{e5.0}{e1.7}, and \numrange{e4.3}{e1.1} times more luminous than the EoR signal on scales of $\SI{0.1}{\per\Mpc} < k < \SI{2}{\per\Mpc}$ in the \numrange{116}{124}, \numrange{146}{154}, and \SIrange{186}{194}{\MHz} frequency bands, respectively. We further calculate the two-dimensional power spectra inside the EoR window and show that the power leaked by Galactic free--free emission can still be significant, as the power ratios can reach about $110\%$--$8000\%$, $30\%$--$2400\%$, and $10\%$--$250\%$ on scales of $\SI{0.5}{\per\Mpc} \lesssim k \lesssim \SI{1}{\per\Mpc}$ in three frequency bands. Therefore, we indicate that the Galactic free--free emission should be carefully treated in future EoR detections. 
\end{abstract}


\section{Introduction}
\label{chap:introduction}

\noindent The Epoch of Reionization (EoR) is a period after the Dark Ages ($z \sim \numrange{30}{200}$) and Cosmic Dawn ($z \sim \numrange{15}{30}$) that lasts from about 300 million to 1 billion years ($z \sim \numrange{5}{15}$; see \citealt{Koopmans15} and references therein), during which the baryonic matter was ionized by the ultraviolet and soft X-ray photons emitted from the first-generation celestial objects (e.g., first stars, and quasars), forming ionized bubbles that gradually grew larger and finally merged. Although the 21 cm emission line of neutral hydrogen (\hi) is regarded as a decisive probe to directly explore the EoR \citep{Fan06,Furlanetto06,Zaroubi13,Furlanetto16}, its detection is currently precluded by the overwhelming foreground contamination. Among various foreground components, the impact of Galactic free--free emission is still poorly understood, and thus it is necessary to create as accurate an all-sky Galactic free--free emission map as possible in the low-frequency (\SIrange{50}{200}{\MHz}) radio band to guide the development of foreground removal techniques.

The Galactic free--free emission cannot be observed directly, because the Galactic synchrotron component dominates the emission at frequencies lower than \SI{10}{\GHz}, while the dust thermal emission becomes overwhelming at frequencies higher than \SI{10}{\GHz}. However, the \Ha emission line (the 3--2 transition of the hydrogen atom at $\lambda$ = \SI{656.28}{\nm}) provides a way to trace the Galactic free--free emission, since they share the same emission measure $\rm{EM}$ $\equiv \int n_{e}^2 dl$ ($n_e$ is the electron density; e.g., \citealt{Marcelin98,McCullough97,Dickinson03,Sims16}). For example, the brightness temperature of the Galactic free--free emission has been related to the \Ha intensity by \citet{Valls98} and \citet{Reynolds00}. \citet{Dickinson03} derived a $95\%$ sky coverage (except the area $|b| < \SI{5}{\degree}$, $l =$ \SI{160}{\degree}--\SI{0}{\degree}--\SI{260}{\degree}) of Galactic free--free emission map at 30 GHz from the absorption-corrected \Ha intensity map based on the Southern H-Alpha Sky Survey Atlas (SHASSA \footnote{\url{http://amundsen.swarthmore.edu/SHASSA}}; \citealt{Gaustad01}) data and Wisconsin H-Alpha Mapper (WHAM \footnote{\url{http://www.astro.wisc.edu/wham}}; \citealt{Haffner03}) data. Note that these results may have been biased since the observed \Ha intensities used in these works are often misunderstood, especially near or at the Galactic plane (e.g., \citealt{Dennison98,Dickinson03}), due to the absorption and scattering. On the other hand, the Galactic free--free emission can also be deduced from the radio recombination lines (RRLs; \citealt{Alves10,Alves12}). For example, a partial-sky ($|b| < \SI{5}{\degree}$, $l =$ \SI{52}{\degree}--\SI{0}{\degree}--\SI{192}{\degree}) Galactic free--free emission map at \SI{1.4}{\GHz} was proposed by \citet{Alves15} based on the observed RRLs map \footnote{\url{http://www.jodrellbank.manchester.ac.uk/research/parkes\_rrl\_survey/}} that was obtained via the \hi observations of \hi Parkes All-Sky Survey (HIPASS; \citealt{Smith96}) and Zone of Avoidance Survey (ZOA; \citealt{Smith98}).

A large part ($\sim 50\%$--$70\%$) of the Galactic \Ha emission is contributed by the recombination process in the \hii regions. Ionized by massive O and B stars, the \hii regions are mostly concentrated on the Galactic plane and become the brightest infrared and radio objects in the spiral arms \citep{Shaver83,Balser11,Paladini04,Anderson14,Anderson15}. Using the Wide-Field Infrared Survey Explorer (WISE) data, \citet{Anderson14} compiled a most complete \hii region catalog that contains 8400 \hii regions (candidates) located at lower latitudes ($|b| \leq \SI{8}{\degree}$) as well as five well-known \hii regions at middle latitudes. The rest ($\sim 30\%$--$50\%$) of the Galactic \Ha emission is attributed to the recombination process in the warm ionized medium (WIM), which is often known as diffuse ionized gas (DIG) in extra-galaxies (e.g., \citealt{Jura79,Miller93,Dove94,Haffner09}). As one important phase of the diffuse interstellar medium (ISM) in our Galaxy, the WIM exhibits a scale height of \SI{\sim 900}{\pc}, a characteristic temperature of \SI{\sim 10,000}{\K}, and a specified volume-filling factor of \numrange{\sim 0.1}{0.4} (\citealt{Wood99}; Wood99 hereafter). How the WIM is ionized and heated, as well as its relationship with other ISM phases, is still unclear \citep{Haffner03,Miller93,Dove00,Haffner99,Reynolds99}. According to the axisymmetric ISM model proposed by \citet{Ferri98b}, which provides the averaged hydrogen number densities for different ISM phases, as the Galactic height ($z$) measured from the Galactic plane increases from $0$ to \SI{5}{\kpc} the hydrogen nucleus number density of the WIM decreases from $1$ to $\rm{10^{-4}~cm^{-3}}$.

It is important to correct the absorption and scattering in the measurement of \Ha intensity in order to accurately obtain the Galactic free--free emission. Although many methods have been proposed to solve this problem (e.g., \citealt{Reynolds90,Dickinson03,Dong11}), there still exists no standard solution owing to the limited dust data \citep{Lehtinen10,Witt10,Brandt12}. On the aspect of the theoretical modeling, Monte Carlo radiative transfer (MCRT) simulation has been proposed to be a useful tool to predict the scattered \Ha intensity and calculate the intrinsic (i.e., without absorption and scattering) \Ha emission in the Milky Way (e.g., Wood99) and extra-galaxies (e.g., \citealt{Schiminovich01,Lee08,Seon09,Jo12,Seon12,Seon13,Seon14,Seon15}). The MCRT algorithm regards the radiation field as a photon flow, in which photons move in the dusty medium \citep{Steinacker13}. For each photon, its starting point, its initial direction of motion, and the place where it interacts with a dust grain are determined in a probabilistic way. Finally, the statistical analysis of the photons can be used to recover the radiation field. The MCRT method offers a variety of obvious advantages, i.e., take scattering into account properly, comparably simple computer programs (just a random number generator together with some basic loops), and easily parallel operation \citep{Noebauer19}. The stochastic fluctuation is unavoidable for the MCRT method, which can be reduced by increasing the number of test particles (e.g., \citealt{Seon15,Murthy16}). Since 1999, Wood and his collaborators have carried out a series of MCRT simulation works to predict the scattering property and the polarization of the \Ha emission in our Galaxy (e.g., Wood99,\citealt{Wood04,Wood05,Wood10,Barnes15}), these simulations and other observations (e.g., \citealt{Reynolds88,Gordon01,Dong11}) show that at higher latitudes the scattered \Ha intensity may contribute up to $\num{\sim 20}\%$ of the total observed \Ha intensity, and its polarization is less than $\num{1}\%$.

In this work, we will focus on the MCRT implementation of Galactic free--free emission and estimate its impacts on the EoR detection by employing the latest configuration of SKA1-Low to incorporate the instrumental effects. The Galactic free--free emission is obtained by performing a three-dimensional (3D) MCRT simulation of Galactic \Ha emission, which comprises the direct and scattered \Ha radiation from \hii regions and the WIM. In the previous MCRT simulations (e.g., Wood99), the \hii regions are treated as simple \enquote{point sources} and the dust is assumed to possess a smooth axisymmetric distribution along the Galactocentric distance. As an improvement we will adopt more realistic models, which are constrained by the multiband observations, to describe the 3D distributions of \hii regions and the dust. To be specific, each \hii region is modeled as a sphere with a radius of $R_{\star}$ and is inserted into our simulation cube according to its Galactic coordinate and the distance to the Sun, which are provided in the WISE \hii catalog. To obtain the dust distribution, we employ the best exponential fitting to the newest observed HI4PI \footnote{\url{http://cdsarc.u-strasbg.fr/viz-bin/qcat?J/A+A/594/A116}} \hi column density map \citep{HI4PI16}. The Thomson scattering of free electrons (see Section \ref{chap:opacity}) is also taken into account by applying a plane-parallel exponential model to describe the distribution of the free electrons. Finally, by analyzing the one-dimensional (1D) and two-dimensional (2D) power spectra and EoR window, we quantitatively evaluate the contamination caused by Galactic free--free emission on the EoR detections.

This paper is organized as follows: In Section \ref{chap:models}, we construct the physical components in the simulation box. In Section \ref{chap:eor-SKA}, we use the \texttt{Simfast21} code to simulate the EoR signal and employ the latest SKA1-Low layout configuration to simulate the SKA \enquote{observed} images. In Section \ref{chap:halpha-Gff}, we present the results of the simulated \Ha intensity and the corresponding Galactic free--free emission, and also evaluate the contamination imposed by Galactic free--free emission on the EoR detection. We discuss the major uncertainties in our simulation and compare our results with the previous works of Wood99 and \citet{Finkbeiner03} (F03 hereafter) in Section \ref{chap:comp-diss}. Finally, we summarize our work in Section \ref{chap:summary}.


\section{Models}
\label{chap:models}

We calculate the full-sky Galactic \Ha intensity by carrying out a 3D MCRT simulation. To determine the scattered \Ha intensities from \hii regions and the WIM, we take the effects of absorption and scattering into account by filling the simulation box with dust and free electrons. The ingredients of the simulation box, \Ha emissivity, clumpy dust, scattering parameters, and radiation transfer algorithm are presented in Sections \ref{chap:grid}--\ref{chap:RT}, respectively. We describe how to relate the Galactic free--free emission with the simulated \Ha intensities in Section \ref{chap:ff-ha}.

\subsection{Simulation Box}
\label{chap:grid}

We carry out the MCRT simulation inside a 3D Cartesian box with \num{1000} $\times$ \num{1000} $\times$ \num{333} cells, which covers a physical size of 30 $\times$ 30 $\times$ 10 $\rm kpc^{3}$ (i.e., the size of each cell is about \num{30} $\times$ \num{30} $\times$ \num{30} $\rm pc^3$), considering that the radial size of Galactic plane and Galactic height $z$ are \SI{\pm 15}{\kpc} and \SI{\pm 5}{\kpc}, respectively. The detector is assumed to be located at (6.5, 15, 5) \si{\kpc} (i.e., the position of Earth) inside the simulation box and can observe the entire sky. By following the work of Wood99, we insert an evacuated region with a radius of \SI{200}{\pc} centered on the Sun to represent the low-density \enquote{Local Bubble} \citep{Cox87}.

\subsection{H$\alpha$ Emissivity}
\label{chap:data}

\begin{figure}[htbp]
	\centering
	\includegraphics[width=0.50\textwidth]{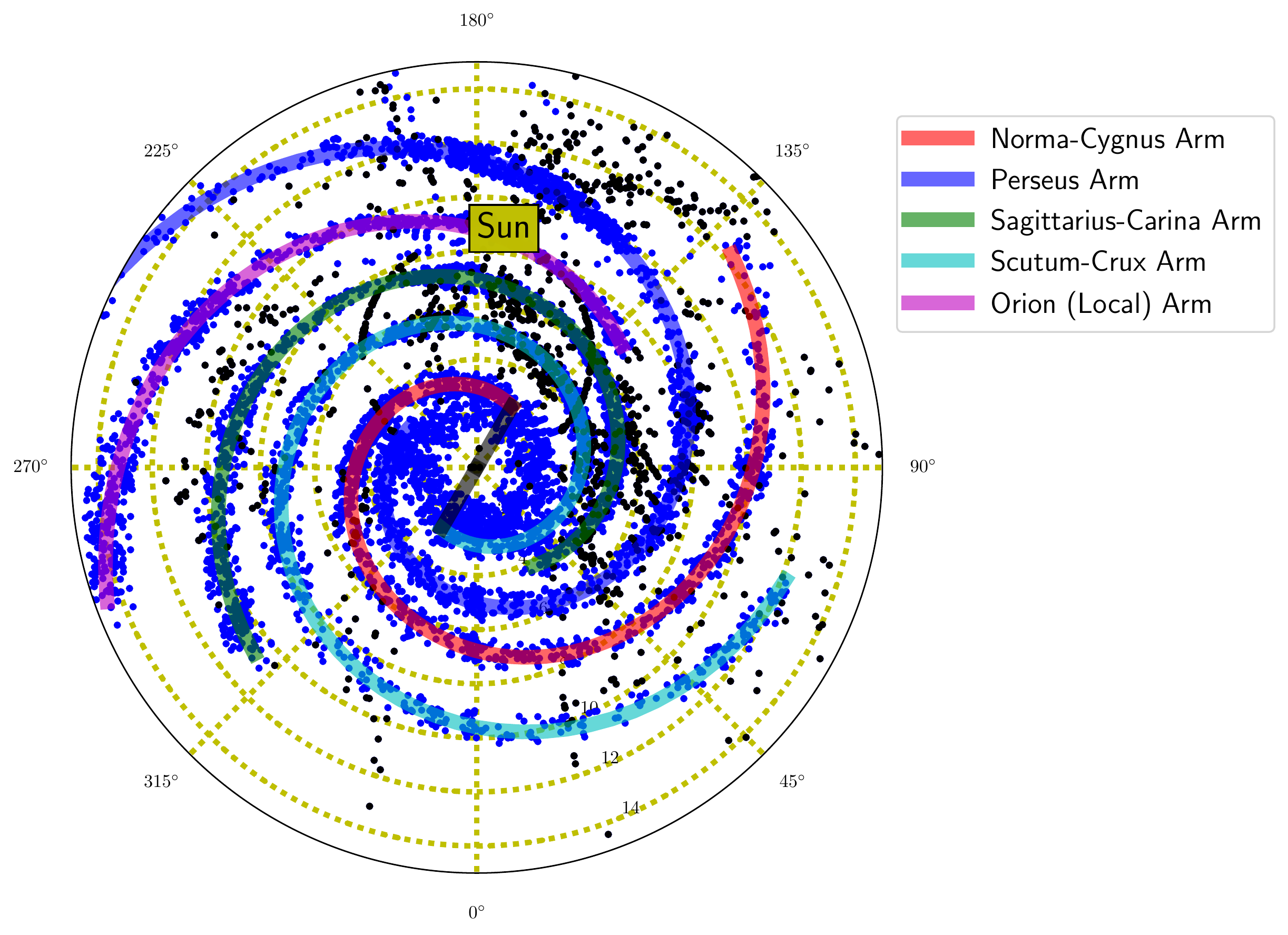}
	\caption{Distribution of \hii regions with the associated Nakanishi16 spiral arms. The yellow square marks the position of the Sun ($180^{\circ}$, 8.5 kpc).}
	\label{fig:arm}
\end{figure}

The \Ha emissivity $\epsilon_{\rm H\alpha}$ is contributed by the emissions from \hii regions and the WIM. It is found that the \Ha emissivity of WIM $\epsilon_{\rm H\alpha}^{\rm WIM}$ is proportional to the square of its hydrogen density (see equation 7 of \citealt{Ferri98b}). By following Wood99's work, we adopt a total WIM emissivity of $10^{52}~\rm{H\alpha~photons~s^{-1}}$. It is expected that the \Ha emissivity near the Galactic plane is dominated by the bright \hii regions. We adopt a total \hii emissivity of $10^{52}~\rm{H\alpha~photons~s^{-1}}$, same as that of the WIM, by following the work of Wood99, since observations show that the volume-averaged ratio of the total \Ha emissivity of \hii regions to that of the WIM is in the range of \numrange{0.47}{0.70} (e.g., \citealt{Veilleux95,Ferguson96}). The \hii regions are placed inside the simulation cube according to their coordinates and the distances to the Sun provided by the WISE \hii catalog\footnote{\url{http://astro.phys.wvu.edu/wise} (version 1.5)} (except for some sources that lack accurate distance data; for more details, see \citealt{Anderson14}). As illustrated in Figure \ref{fig:arm}, the \num{1546} \hii regions possessing known radii and distances to the Sun are shown with black circles, and the other \num{6859} sources that lack the distances information are shown with blue circles, for which, by following Wood99's work, we assign random positions by placing sources randomly in the molecular ring \citep{Ferri98b} and the spiral arms \citep{Nakanishi16}. We also display the spiral arms of our Galaxy given by \citet{Nakanishi16} in Figure \ref{fig:arm} for comparison.

\begin{figure}[htbp]
	\centering
	\includegraphics[width=0.50\textwidth]{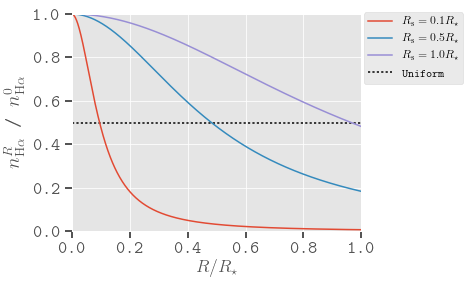}
	\caption{Four types of \hii region model, i.e., variations of \Ha photon density $n_{\rm H\alpha}^{R} / n_{\rm H\alpha}^{0}$ with the radius $R / R_{\rm s}$, which are labeled as $R_{\rm s} = 0.1 R_{\star}$ (blue solid line), $R_{\rm s} = 0.5 R_{\star}$ (orange solid line), $R_{\rm s} = 1.0 R_{\star}$ (green solid line), and \enquote{Uniform} (black dashed line).}
	\label{fig:hii-models}
\end{figure}

To quantify the spatial distribution of \Ha photons in each \hii region, we adopt a $\beta$ profile $n_{\rm H\alpha}^{R} = n_{\rm H\alpha}^{0} (1 + (\frac{R}{R_{\rm s}})^{2})^{-3\beta/2}$ ($\beta$ \hii region model hereafter; \citealt{Cavaliere76}), where $R$ is the distance measured from the \hii region center, $\beta = 0.7$ is the slope parameter, $R_{\rm s}$ is the scale radius ($R_{\rm s} / R_{\star} = 0.1$, $0.5$, and $1.0$ have been tested, where $R_{\star}$ is the \hii region's radius provided by the WISE \hii catalog), and $n_{\rm H\alpha}^{0}$ is the density of \Ha photons at the center of the \hii region, whose value is about several times of $10^{49}~\rm{H\alpha~photons~s^{-1}}$ (the typical value of Orion Nebula; Wood99), which can be determined from the total \Ha photons in each \hii region. For comparison, we also test a uniform model of \hii regions, i.e., the distribution of \Ha photons in each \hii region is uniform. The four types of \hii region model, i.e., $\beta$ cases with $R_{\rm s} = 0.1 R_{\star}$, $0.5 R_{\star}$, $1.0 R_{\star}$, and \enquote{Uniform} case, are shown in Figure \ref{fig:hii-models}. 

\subsection{Distributions of Dust and Free Electrons}
\label{chap:opacity}

We insert the dust and free electrons into our simulation cube to calculate the scattered \Ha emission. Given that the dust optical depth at the \Ha wavelength can be calculated via the \hi column density \citep{Bohlin78}, we employ an exponential fitting $N_{\rm H{\sc I}}^{D}$ = $N_{\rm H{\sc I}}^{0}~e^{- D / 125 \rm pc}$ to derive the 3D distribution of \hi column density, where \SI{125}{\pc} is the scale length \citep{Marshall06}, $D$ is the distance to the Sun, and $N_{\rm H{\sc I}}^{0}$ is the \hi column density at the Galactic plane that can be derived from the 2D HI4PI \hi column density map. The HI4PI is an all-sky 2D \hi column density map, which is obtained from the observed data of the Effelsberg-Bonn \hi Survey and Galactic All-Sky Survey \citep{HI4PI16}. A plane-parallel exponential model of free electrons $n_{\rm{e}}(z) = n_{\rm {e,0}}~e^{-|z|/h}$ is applied to obtain the 3D distribution of free electrons, where $n_{\rm e,0}$ is the free electron density at the Galactic plane and $h$ is the scale height of free electron density \citep{Schnitzeler12}. We adopt $n_{\rm{e,0}} = 0.0165~\rm{cm^{-3}}$ and $h = \SI{1.45}{\kpc}$, which are calculated based on the table 3 of \citet{Schnitzeler12}.

\subsection{Scattering Parameters}
\label{chap:scattering}

The scattered \Ha emission is simulated by labeling each \Ha photon and tracing its traveling routes in our simulation. To calculate the scattered \Ha intensity, we employ the Henyey--Greenstein (HG) phase function \citep{HG41}

\begin{equation}
\label{equ1:HG}
{\rm {HG}(\theta)} = \frac{1}{4\pi}~\frac{1 - g^2}{[1 + g^2 - 2g~{\rm cos}(\theta)]^{3/2}},
\end{equation}

\noindent where $\theta$ (in the range of [0, $\pi$]) is the scattering angle, so that $\theta = 0$ corresponds to forward scattering, $\theta = \pi$ means back scattering, and $g$ ($g \equiv <{\rm cos}(\theta)>$) is the phase function asymmetry factor, and $g > 0$ indicates forward scattering predominance. Three typical sets of scattering parameters, i.e., $g = 0.44$, $0.50$, and $0.55$, have been tested in our simulation (see also Table \ref{Tab1:simulation}; \citealt{Mathis77,Weingartner01}).

\subsection{Monte Carlo Radiative Transfer}
\label{chap:RT}
\begin{table}
	\caption{Model Parameters in Our Simulation}
	\label{Tab1:simulation}
	\begin{center}
		\begin{tabular}{lcccccccccccccc}
			\hline
			\hline
			Label  & $a$& $g$& \hii Model \\
			\hline
			(a) &  \num{0.50} & \num{0.44}  & $R_{\rm s} = 0.1 R_{\star}$ \\
			(b) &  \num{0.50} & \num{0.44}  & $R_{\rm s} = 0.5 R_{\star}$\\
			(c) &  \num{0.50} & \num{0.44}  & $R_{\rm s} = 1.0 R_{\star}$\\
			(d) &  \num{0.50} & \num{0.44}  & Uniform\\
			(e) &  \num{0.67} & \num{0.50}  & $R_{\rm s} = 0.1 R_{\star}$\\
			(f) &  \num{0.67} & \num{0.50}  & $R_{\rm s} = 0.5 R_{\star}$ \\
			(g) &  \num{0.67} & \num{0.50}  & $R_{\rm s} = 1.0 R_{\star}$\\
			(h) &  \num{0.67} & \num{0.50}  & Uniform\\
			(i) &  \num{0.77} & \num{0.55}  & $R_{\rm s} = 0.1 R_{\star}$\\
			(j) &  \num{0.77} & \num{0.55}  & $R_{\rm s} = 0.5 R_{\star}$\\
			(k) &  \num{0.77} & \num{0.55}  & $R_{\rm s} = 1.0 R_{\star}$\\
			(l) &  \num{0.77} & \num{0.55}  & Uniform\\
			\hline
			(A) &  ---  & --- &  $R_{\rm s} = 0.1 R_{\star}$\\
			(B) &  ---  & --- & $R_{\rm s} = 0.5 R_{\star}$\\
			(C) &  ---  & --- & $R_{\rm s} = 1.0 R_{\star}$\\
			(D) &  ---  & --- & Uniform\\
			\hline
			\hline
		\end{tabular}
	\end{center}
\end{table}

The design of our simulation code is similar to other MCRT programs in predicting the direct and scattered \Ha intensities (e.g., Wood99; \citealt{Gordon01,Barnes15}). We will present the code flow in our simulation by tracing the motion of a single \Ha photon.

(\romannumeral1) An \Ha photon is emitted from either \hii regions or the WIM according to their weighted distributions. Each photon begins with initial effective unit weight and is forced to send a fraction of weight $W_{\rm{direct}}$ to the detector  

\begin{equation}
\label{equ:direct}
W_{\rm{direct}} = e^{-\tau} / 4\pi d^2,
\end{equation}

\noindent where $d$ is the distance from the point of emitter to the detector; $\tau$ is the optical depth of \Ha emission along the distance of $d$, i.e., $\tau$ = $\int_{0}^{d} (N_{\rm HI} \sigma_{\rm H\alpha} + N_{{\rm e}} \sigma_{\rm t})dl$; $N_{\rm HI}$ and $N_{{\rm e}}$ are the \hi column density and free electron column density, respectively; $\sigma_{\rm H\alpha}$ = $3.801 \times 10^{-22}~\rm{cm^2}$ is the scattering cross section at \Ha wavelength \citep{Draine03}; and $\sigma_{\rm t}$ = $6.652 \times 10^{-25}~\rm{cm^2}$ is the Thomson cross section.

(\romannumeral2) Next, two random numbers are generated to determine the direction of \Ha photon motion, one for theta (in the range of [0, $\pi$], measured from the $z$-axis of our simulation cube) and the other for phi (in the range of [0, 2$\pi$]). To calculate the scattered \Ha intensity, by following \citet{Murthy16}, a third random number $\xi$ is generated from a uniform distribution [0, 1] to determine a predetermined optical depth $\tau_{\rm pre}$, which is sampled from --log$(\xi)$. Then, the scattering location is determined by following the \Ha photon's motion until the cumulative optical depth $\tau_{\rm cum}$ along the path equals the $\tau_{\rm pre}$. If this location is inside the simulation box, we apply the \enquote{peel-off} strategy to calculate the scattered weight received by the detector (e.g., Wood99; \citealt{Yusef84})

\begin{equation}
\label{equ:scatter}
W_{\rm scatter}^{N} = a~~W_{\rm rest}^{N-1}~~(1 - e^{-\tau_{\rm pre}^{N}})~~ e^{-\tau_{\rm scatter}^{N}}~~ {\rm{HG}}(\theta)~~/~~ d_N^2,
\end{equation}

\noindent where $a$ is the reflectivity or albedo ($a = 0.50$, $0.67$, and $0.77$ are adopted; see also Table \ref{Tab1:simulation}; \citealt{Weingartner01,Murthy16}), $N$ is the scattering count, $W_{\rm scatter}^{N}$ is the $N$ times scattering weight, $W_{\rm rest}^{N-1}$ is the rest weight after $N-1$ times scattering, $\tau_{\rm pre}^{N}$ is the $N$ times predetermined scattering optical depth, $\tau_{scatter}^N$ is the $N$ times scattering optical depth of \Ha emission obtained by $\tau_{\rm scatter}^N$ = $\int_{0}^{d_N} (N_{\rm HI} \sigma_{\rm dust} + N_{{\rm e}} \sigma_{\rm t})dl$ ($d_N$ is the distance of $N$ times scattering point to the detector), the scattering phase function ${\rm{HG}}(\theta)$ is given in Equation \ref{equ1:HG}, and the scattering angle $\theta$ is obtained by ${\rm arccos}$$(\mathbf{v} \cdot \mathbf{l} / |\mathbf{v}||\mathbf{l}|)$, where $\mathbf{v}$ and $\mathbf{l}$ are the vectors along the direction of motion and the direction toward the detector, respectively. 

Then, a new scattering direction (once the scattering process begins, the theta should be weighted by the ${\rm HG}(\theta)$ function) and a new predetermined optical depth are generated, and the \Ha photon is tracked until it exits the box. In actual simulations, we can set a threshold (e.g., $10^{-20}$; \citealt{Steinacker13}) for the effective weight of \Ha photons or the maximum scattering number (e.g., $1000$; \citealt{Murthy16}) to terminate the scattering process.

(\romannumeral3) The total \Ha intensity is the sum of the weights of the direct (Equation \ref{equ:direct}) and multiple scattered (Equation \ref{equ:scatter}) photons. The received \Ha photons are then used to construct the \Ha intensity map using the Hierarchical Equal Area isoLatitude Pixelization (HEALPix) \footnote{\url{http://healpix.sourceforge.net/}} tessellation scheme with $N_{\rm{side}}$ = $1024$ (pixel size $\simeq \SI{3.44}{\arcminute}$; \citealt{Goski05}).

\begin{figure*}[htbp]
	\centering
	\includegraphics[width=1.0\textwidth]{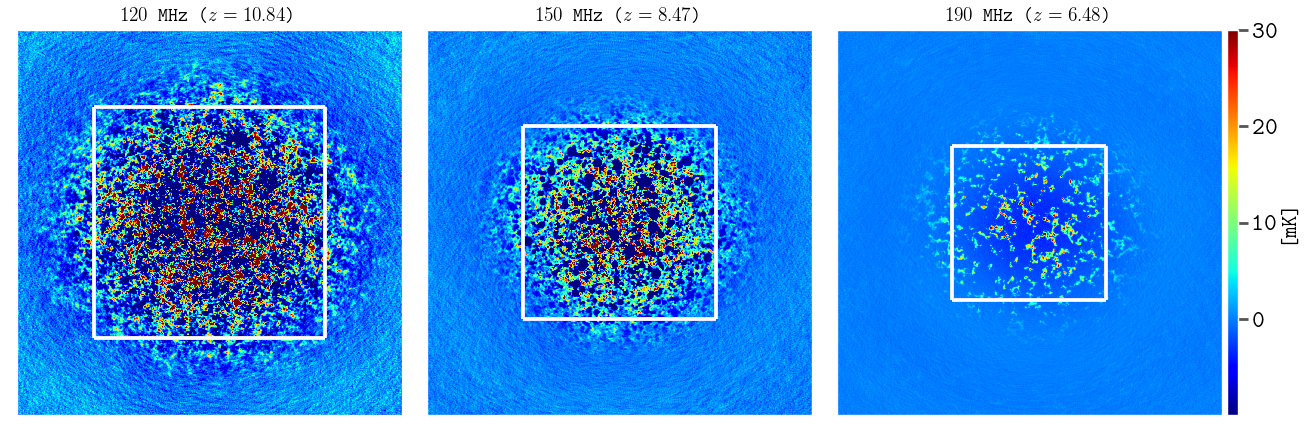}
	\caption{Brightness temperatures of the SKA \enquote{observed} EoR signal at \numlist{120; 150; 190} \si{\MHz}. The sky coverage is \SI[product-units=repeat]{10 x 10}{\degree}, and the color bar is in units of \si{\mK}. The white squares mark the \SI[product-units=repeat]{6 x 6}{\degree}, \SI[product-units=repeat]{5 x 5}{\degree}, and \SI[product-units=repeat]{4 x 4}{\degree} central regions at three frequencies.} 
	\label{fig:21cm}
\end{figure*}

\subsection{Derivation of Free--Free Emission}
\label{chap:ff-ha}

The received \Ha intensity depends on whether the emitting medium is optically thin (case A) or optically thick (case B), and it is found that the case B is satisfied in the study of Galactic \Ha emission \citep{Osterbrock89,Dickinson03}. For case B, \citet{Valls98} proposed an analytical expression to describe the relation between the observed \Ha intensity $I_{\rm H\alpha}(\mathbf{r})$ and the emission measure ${\rm EM}(\mathbf{r})$

\begin{equation}
\label{equ:EM}
{\rm EM}(\mathbf{r}) = 2.561~~ T_{4}^{1.017}(\mathbf{r}) ~~10^{0.029/T_{4}(\mathbf{r})}~~ I_{\rm H\alpha}(\mathbf{r}),
\end{equation}

\noindent where $T_{4}(\mathbf{r})$ = $T_{\rm e}(\mathbf{r}) / 10^4$ ($\mathbf{r}$ is the 2D position) is the electron temperature in units of \SI{e4}{\K}, $I_{\rm H\alpha}(\mathbf{r})$ is in units of Rayleigh (${\rm R}$) \footnote{1 Rayleigh $\rm{(R)} \equiv 10^{6}/4\pi~$${\rm photons~s^{-1}~cm^{-2}~sr^{-1}}$ $\equiv 2.41 \times 10^{-7} \rm{erg~cm^{-2}~s^{-1}~sr^{-1}}$}, and the ${\rm EM}(\mathbf{r})$ is in units of $\rm cm^{-6}~pc$. Using the emission measure derived in Equation \ref{equ:EM}, we can calculate the optical depth of Galactic free--free emission $\tau_{c}(\mathbf{r})$ as

\begin{equation}
\label{equ:tau-Draine}
\tau_{c}(\mathbf{r}) = 0.05468~~g(\mathbf{r})~~T_{\rm e}(\mathbf{r})^{-3/2}~~ \nu_{9}^{-2}~~ {\rm EM}(\mathbf{r}),
\end{equation}

\noindent where $\nu_{9}$ = $\nu / 10^9$ \si{\Hz} is the frequency in units of \si{\GHz}, and $g(\mathbf{r})$ is the gaunt factor given by

\begin{equation}
\label{equ:g-Draine}
g(\mathbf{r}) = {\rm log} \{{\rm exp[5.960 - \sqrt{3}}/\pi~{\rm log}(\nu_{9}~T_{4}(\mathbf{r})^{-3/2})] + \rm e\},
\end{equation}	

\noindent where $\rm e \simeq 2.71828...$ is the natural constant \citep{Draine11}. The above three equations are valid in the \SI{100}{\MHz}--\SI{100}{\GHz} frequency bands \citep{Dickinson03}, which are often employed to deduce the brightness temperature of Galactic free--free emission $T_{\rm b}^{\rm {Gff}}(\mathbf{r})$

\begin{equation}
\label{equ:tb}
T_{\rm b}^{\rm {Gff}}(\mathbf{r}) = T_{\rm e}(\mathbf{r})~~ [1 - {\rm e}^{- \tau_{\rm c}(\mathbf{r})}].
\end{equation}


\section{SKA Observation and EoR Signal}
\label{chap:eor-SKA}

In order to incorporate the instrumental effects of radio interferometers, we have employed the latest SKA1-Low layout configuration \footnote{\raggedright%
	SKA1-Low Configuration Coordinates:
	\url{https://astronomers.skatelescope.org/wp-content/uploads/2016/09/SKA-TEL-SKO-0000422_02_SKA1_LowConfigurationCoordinates-1.pdf}
	(released on 2016 May 31)
} to simulate the SKA \enquote{observed} images. The SKA1-Low interferometer layout includes $512$ stations, with $224$ stations randomly distributing within the \enquote{core} region (\SI{1000}{\meter} in diameter), and others scattering in \enquote{cluster} regions, which form three spiral arms up to a radius of \SI{\sim 35}{\km}. Each station includes $256$ antennas, which are randomly distributed in a circular region of \SI{35}{\meter} in diameter with a minimum separation of $d_{\rm min}$ = \SI{1.5}{\meter} (e.g., \citealt{Mort17}).

We choose the sky maps centered at ($\rm R.A.$, $\rm Dec.$) = (\SI{0}{\degree}, \SI{-30}{\degree}) with a sky coverage of \SI[product-units=repeat]{10 x 10}{\degree}, which is located at a high galactic latitude ($b = \SI{-78}{\degree}$) and is expected to be an appropriate choice for this study. Moreover, this region passes through the zenith of the SKA1-Low telescope and is an ideal choice to simulate the SKA observation. We use the \texttt{OSKAR} \footnote{OSKAR: \url{https://github.com/OxfordSKA/OSKAR (version 2.7.0)}} \citep{Mort10} simulator to perform SKA observations for 6 hr to obtain the visibility data. The \texttt{WSClean} imager \citep{Offringa14} is employed to image the simulated visibility data using Briggs weighting with a zero robustness \citep{Briggs95,Li19}. To avoid the problem of insufficient CLEAN in the marginal regions, we crop the created images and choose to keep their central regions of \SI[product-units=repeat]{6 x 6}{\degree}, \SI[product-units=repeat]{5 x 5}{\degree}, and \SI[product-units=repeat]{4 x 4}{\degree} in the \numrange{116}{124}, \numrange{146}{154}, and \SIrange{186}{194}{\MHz} frequency bands, respectively, given that the telescope's field of view (FOV) is inversely proportional to the observing frequency (see the example maps in Figure \ref{fig:21cm}). For each frequency band, the CLEAN algorithm with joined-channel deconvolution is adopted to create the foreground cube in order to ensure the spectral smoothness \citep{Offringa17}. We directly use the dirty image for the EoR signal, because the CLEAN algorithm does not work well for such faint diffuse emission.

The seminumerical code \texttt{Simfast21} \footnote{\url{https://github.com/mariogrs/Simfast21}} (\citealt{Santos10,Hassan16}) is employed to simulate the brightness temperatures of the 21 cm signal during the EoR by following our previous work (for more details about the \texttt{Simfast21} simulation, see \citealt{Lian20}). To construct the EoR signal cube, we assume a $\Lambda$CDM cosmology with parameters of $\Omega _{m} = \Omega_{dm} + \Omega_{b} = 0.3089$, $\Omega_{b} = 0.0486$, $\Omega_{\Lambda} = 0.6911$, Hubble constant $H_0 = \SI{67.74}{\km\per\second\per\Mpc}$, power spectrum index $n_{\rm s} = 0.9667$, and the normalization $\sigma_8 = 0.8159$ \citep{PlanckXIII16}. We initialize the \texttt{Simfast21} at $z_{\rm i} = 100$ on a $1024^3$ box with physical dimensions of $1.6^3$ comoving $\rm Gpc^{3}$, which corresponds to a field of $\theta_x$ = $\theta_y$ $\approx$ \SI{9.88}{\degree}, a pixel resolution of $\Delta\theta_x$ = $\Delta\theta_y$ $\approx$ \SI{0.58}{\arcminute}, and a frequency depth of $\Delta\nu$ $\approx$ \SI{92.95}{\MHz}. We then utilize the method of \citet{Mellema06} to create the observable \enquote{light-cone} object using the outputs (the so-called \enquote{coeval cubes}) of \texttt{Simfast21}. From the derived \enquote{light-core} object, we extract three subsets with a channel width of \SI{160}{\kHz} and construct our final tiled data cube with dimensions of ($\theta_x$, $\theta_y$, $\Delta\nu$) = (\SI{10}{\degree}, \SI{10}{\degree}, \SI{8}{\MHz}) in the \numrange{116}{124}, \numrange{146}{154}, and \SIrange{186}{194}{\MHz} frequency bands, among which each image is performed for the SKA \enquote{observed} simulation.  We present the example SKA \enquote{observed} EoR images at \num{120} ($z = 10.84$), \num{150} ($z = 8.47$), and \SI{190}{\MHz} ($z = 6.48$) in Figure \ref{fig:21cm}. The rms brightness temperatures of $\delta T_{\rm b}^{21}$ are \numlist{21.03; 13.12; 4.43} \si{\mK} inside the central regions of \SI[product-units=repeat]{6 x 6}{\degree}, \SI[product-units=repeat]{5 x 5}{\degree}, and \SI[product-units=repeat]{4 x 4}{\degree} at \numlist{120; 150; 190} \si{\MHz}, respectively.


\section{Results}
\label{chap:halpha-Gff}

\begin{figure*}[htbp]
	\centering
	\includegraphics[width=0.325\textwidth]{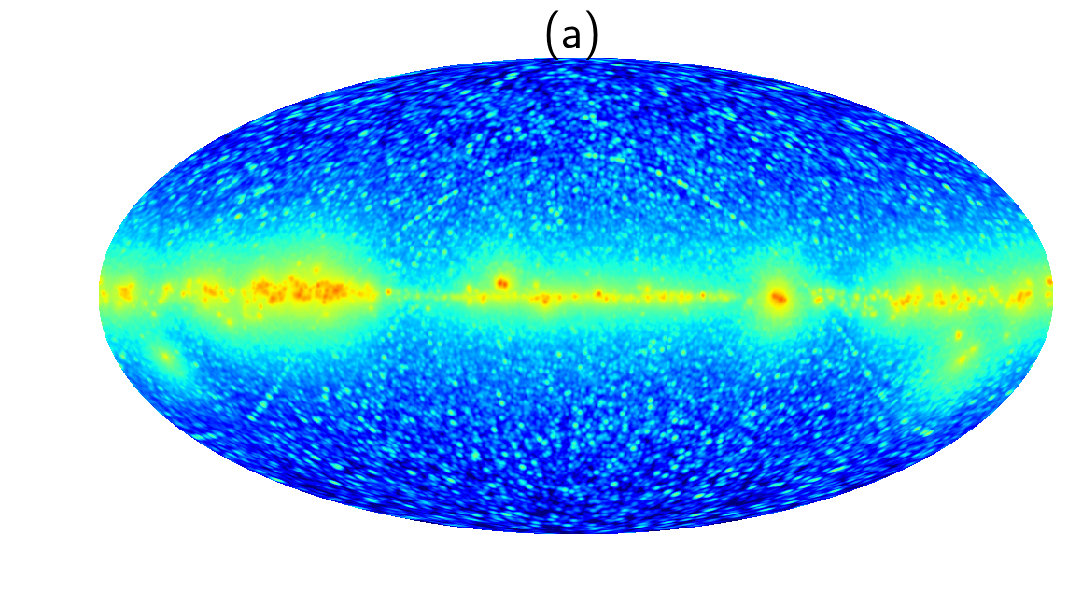}
	\includegraphics[width=0.325\textwidth]{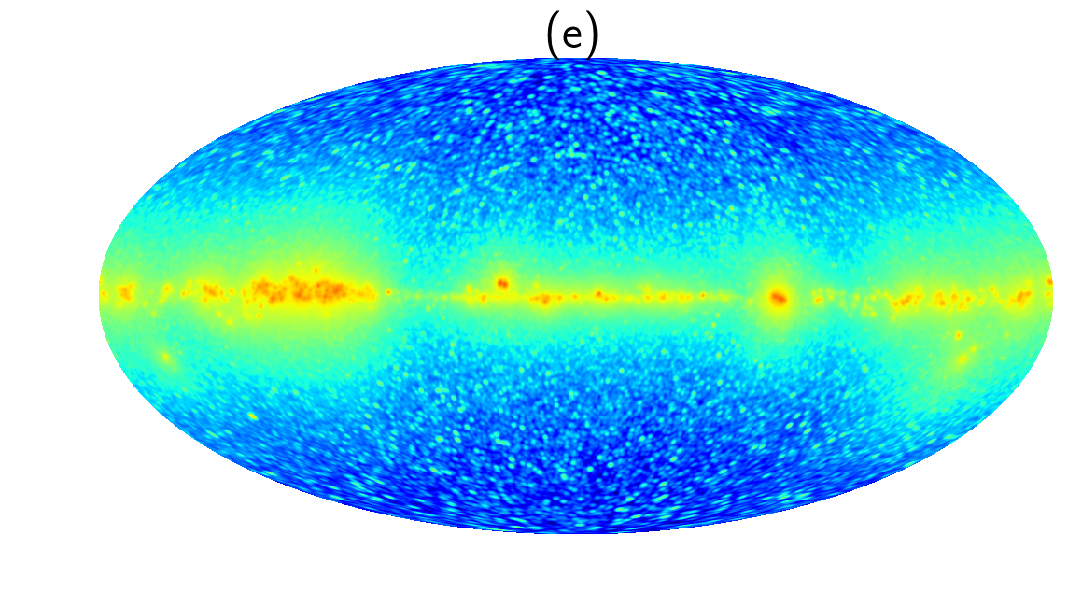}
	\includegraphics[width=0.325\textwidth]{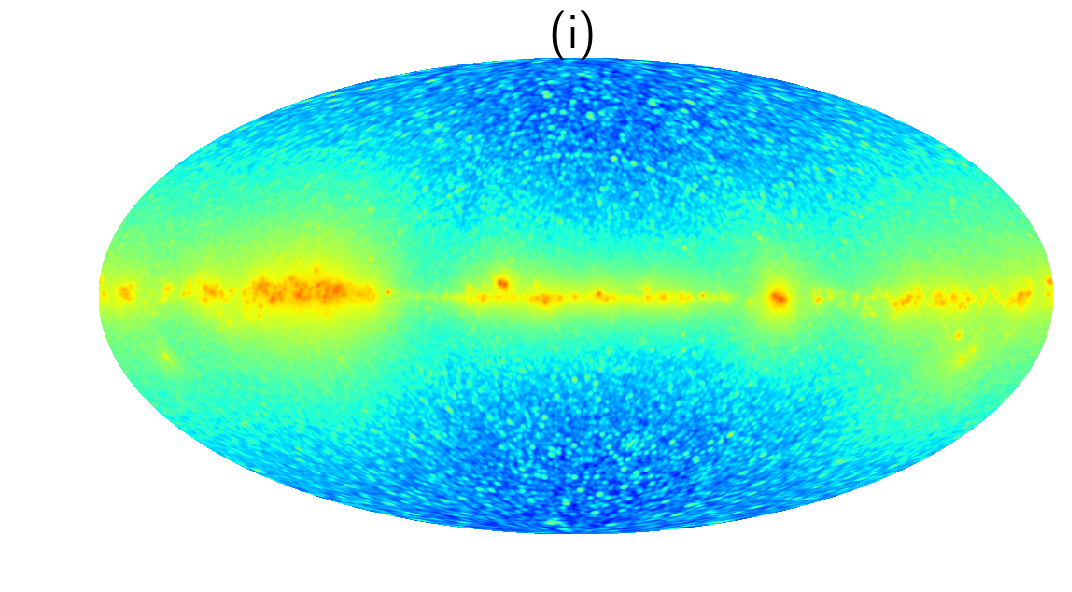}
	\includegraphics[width=0.325\textwidth]{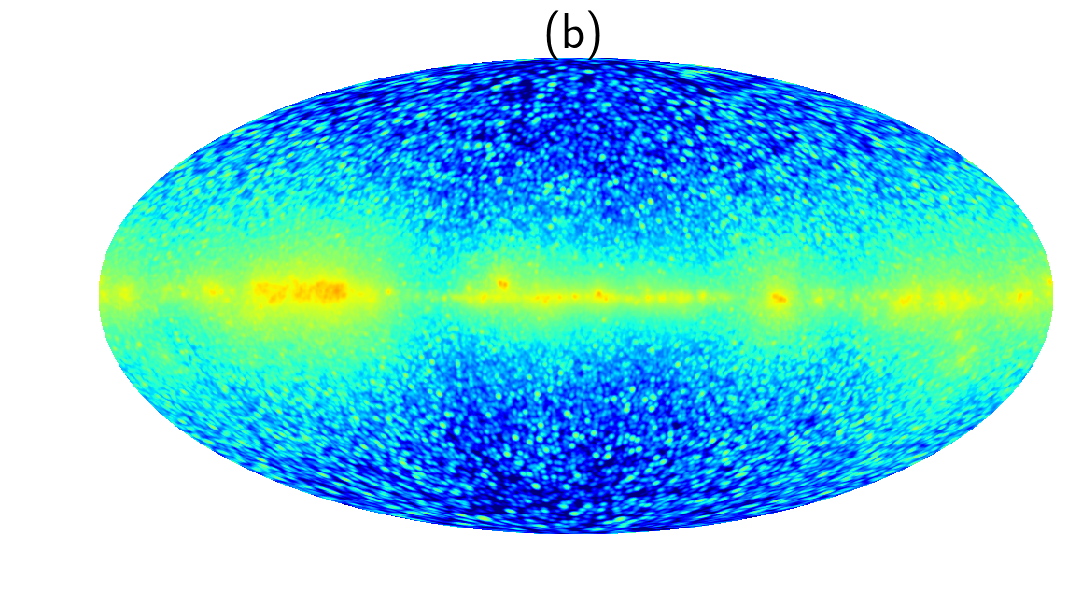}
	\includegraphics[width=0.325\textwidth]{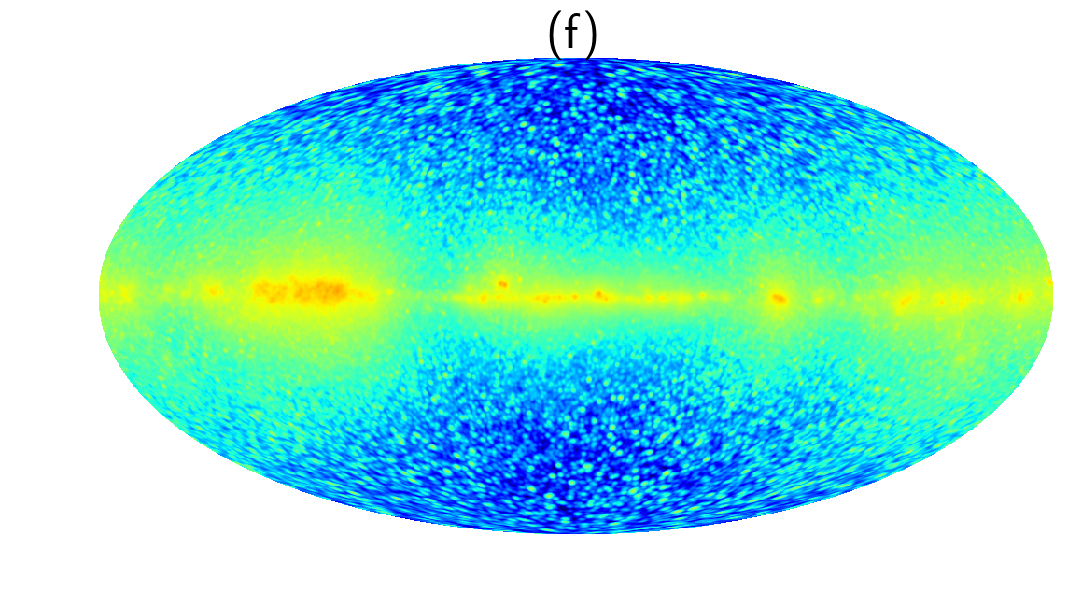}
	\includegraphics[width=0.325\textwidth]{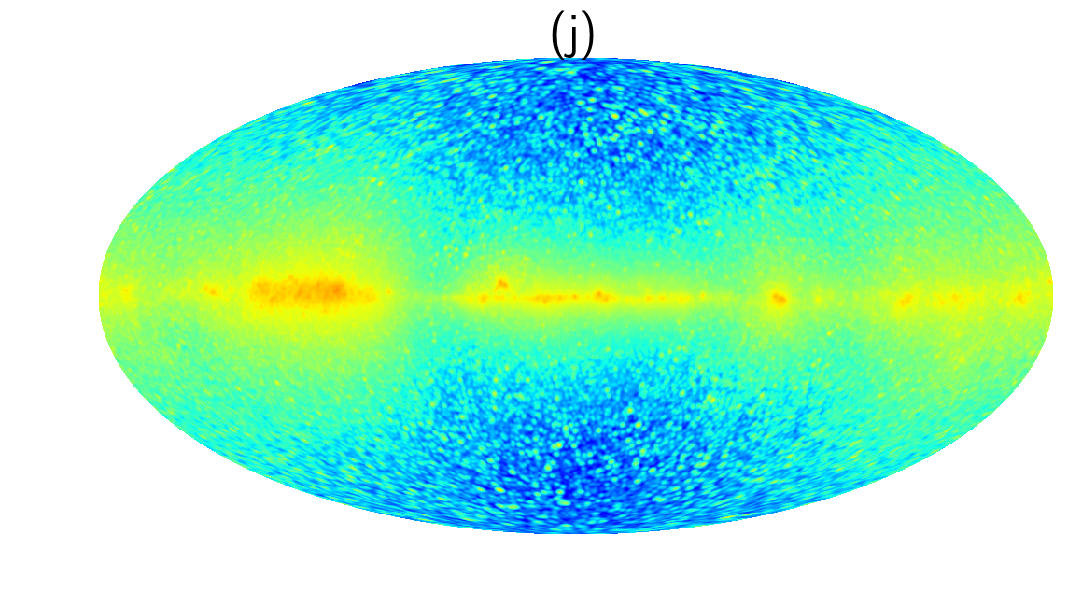}
	\includegraphics[width=0.325\textwidth]{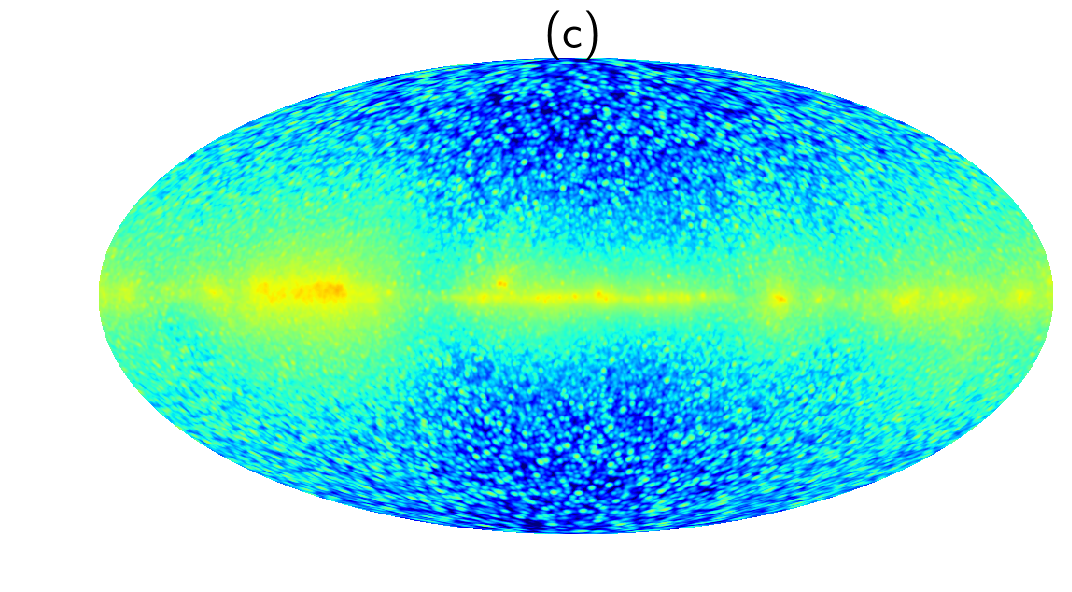}
	\includegraphics[width=0.325\textwidth]{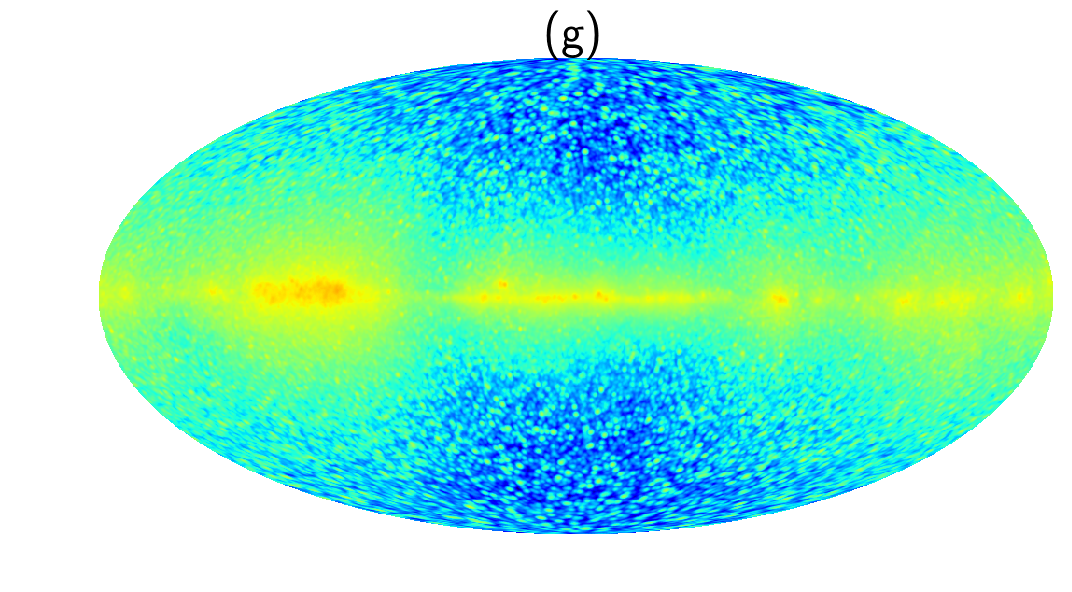}
	\includegraphics[width=0.325\textwidth]{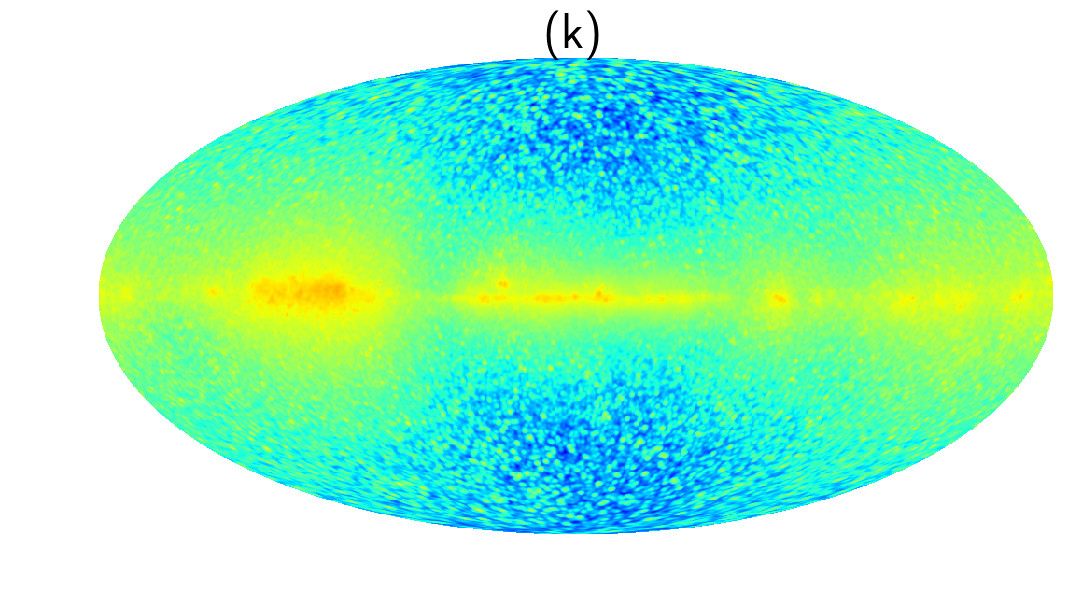}
	\includegraphics[width=0.325\textwidth]{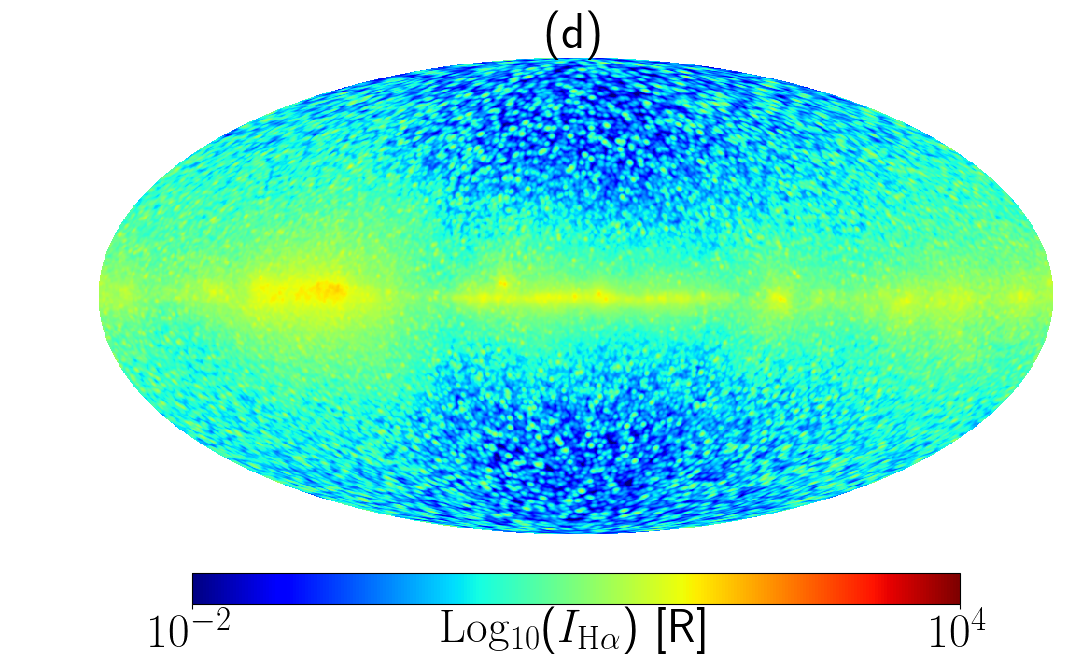}
	\includegraphics[width=0.325\textwidth]{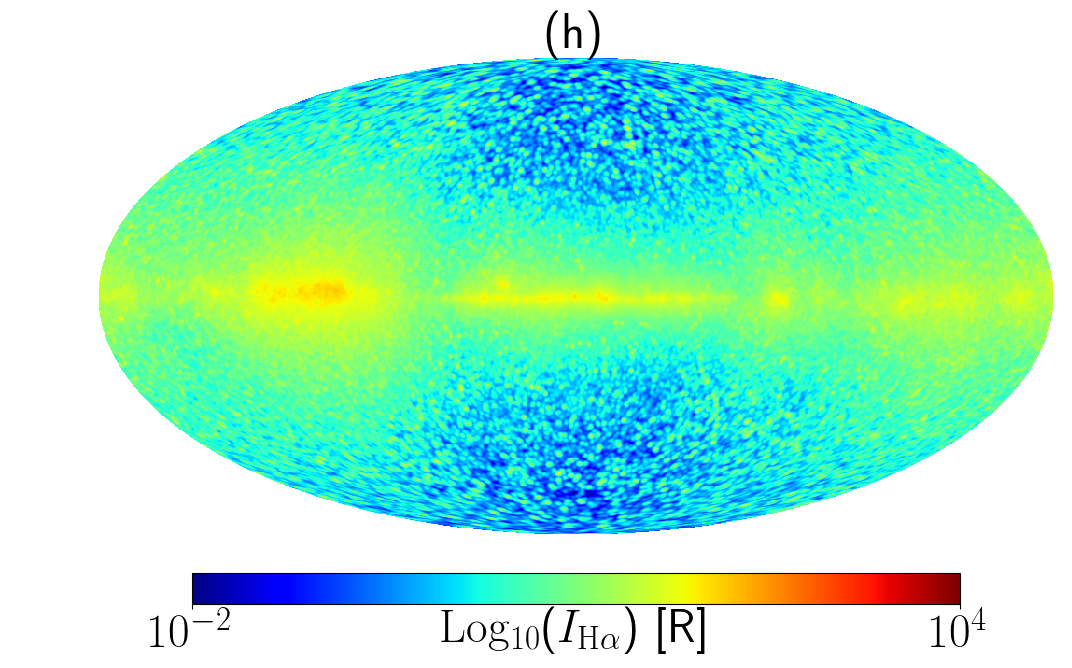}
	\includegraphics[width=0.325\textwidth]{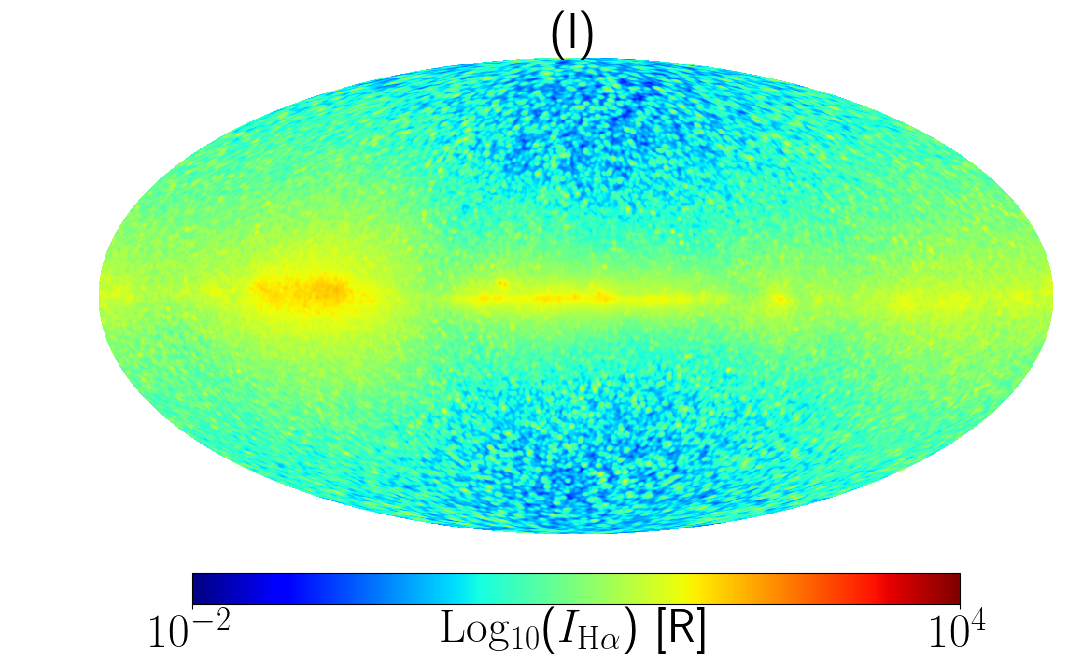}	
	\caption{All-sky Mollweide projections ($N_{\rm side} = 1024$) of the simulated total H$\alpha$ intensity maps in Galactic coordinates. The Galactic center $(l,~b)$ = (\SI{0}{\degree}, \SI{0}{\degree}) is at the center of each figure, and the Orion is on the right below the Galactic plane. All panels show the mean values of $50$ simulation runs and share the same logarithmic scale {\bfseries{in units of $\rm R$.}}}  
	\label{fig:halpha-dust}
\end{figure*}
We present the simulated \Ha intensity maps including $12$ cases of direct and scattered \Ha intensities and $4$ cases of intrinsic \Ha intensities, in Section \ref{chap:halpha}. Meanwhile, we derive the corresponding Galactic free--free emission maps from the simulated \Ha intensity maps in Section \ref{chap:free--free}. Furthermore, by analyzing the 1D and 2D power spectra, we have quantitatively evaluated the contamination imposed by the Galactic free--free emission on the EoR detection in Section \ref{chap:Gff2EoR}.

\subsection{\Ha Intensity Maps}
\label{chap:halpha}

\subsubsection{Direct and Scattered \Ha Intensities}
\label{chap:dusty-halpha}

We perform $12$ cases of simulations with diverse model parameters that are listed in Table \ref{Tab1:simulation}. Each case is repeated $50$ times to estimate the mean and the standard deviation ($1\sigma$) of the simulated total \Ha intensity ($I_{\rm H\alpha}^{\rm {tot}}$) maps. The final mean $I_{\rm H\alpha}^{\rm {tot}}$ maps are Gaussian filtered and smoothed to \SI{1}{\degree} to reduce the Poisson noise, as shown in Figure \ref{fig:halpha-dust}. We compare the results obtained with $\beta$ \hii region models with the uniform \hii region model and find that when the uniform \hii region model is applied the highest \Ha intensity is obtained because fewer \Ha photons concentrated on the Galactic plane, where the absorption is severest. We further compare the $I_{\rm H\alpha}^{\rm {tot}}$ maps with different scattering parameters and confirm that increasing $a$ and $g$ will increase the \Ha intensity. The averaged $I_{\rm H\alpha}^{\rm {tot}}$ of the whole sky for 12 cases and their corresponding standard deviations are listed in Table \ref{Tab2:halpha-sff}, with the values of $4.43 (\pm 0.30)$--$11.24 (\pm 0.74)$ $\rm R$.

\begin{figure*}[htbp]
	\centering
	\includegraphics[width=1.0\textwidth]{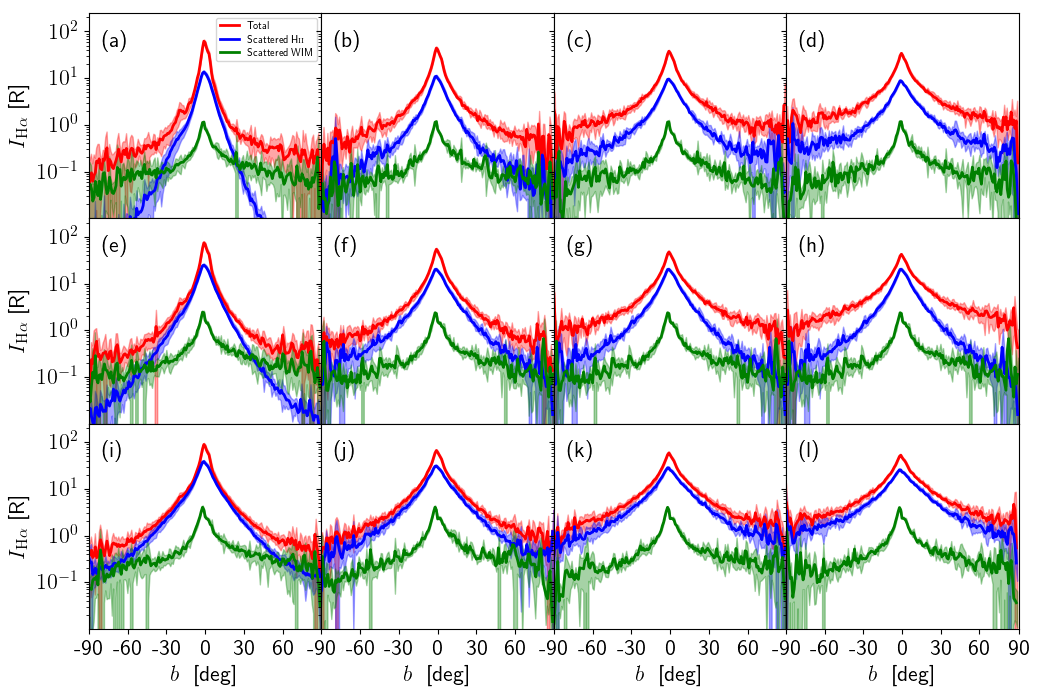}
	\caption{Latitudinal distributions of the simulated total \Ha intensities, the scattered \Ha intensities from \hii regions, and the scattered \Ha intensities from WIM. The solid lines and shaded regions show the mean values and the corresponding $1\sigma$ uncertainties estimated from $50$ simulation runs, respectively.}
	\label{fig:halpha-plot}
\end{figure*}

\begin{figure*}[htbp]
	\centering
	\includegraphics[width=1.0\textwidth]{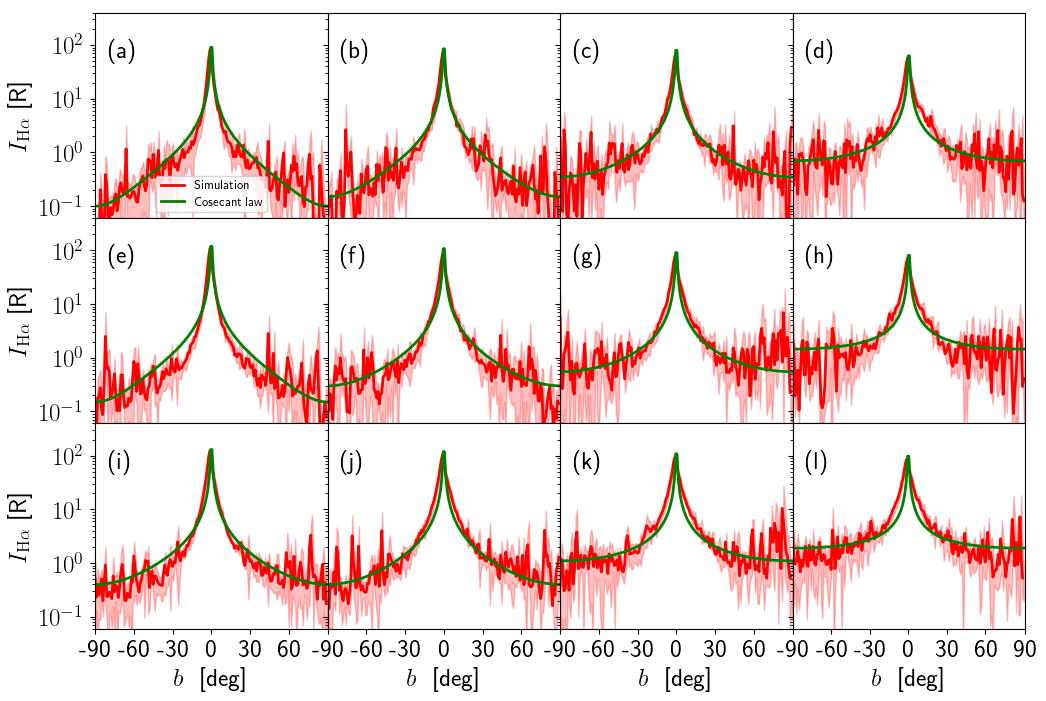}
	\caption{Cosecant fitting profiles (green solid lines) for \SI{30}{\degree}-wide latitudinal cuts of the simulated total \Ha intensities. The red solid lines and red shaded regions illustrate the mean and the corresponding $1\sigma$ uncertainties of $50$ simulation runs, respectively.}
	\label{fig:halpha-fitting}
\end{figure*}

We then present the latitudinal distributions of the mean and $1\sigma$ uncertainty of the simulated \Ha intensities in Figure \ref{fig:halpha-plot}. We find that at lower latitudes ($|b| \lesssim \SI{8}{\degree}$) the $I_{\rm H\alpha}^{\rm {tot}}$ obtained with $R_{\rm s} = 0.5 R_{\star}$, $R_{\rm s} = 1.0 R_{\star}$, and a uniform \hii region model show lower values than those obtained with $R_{\rm s} = 0.1 R_{\star}$ by about $15\%$, $25\%$, and $40\%$, respectively. However, at middle and higher latitudes ($|b| \gtrsim \SI{8}{\degree}$) a contrary tendency is found, as the corresponding simulated \Ha intensities become about \numlist{2; 3; 5} times higher than those obtained with $R_{\rm s} = 0.1 R_{\star}$. We also confirm that increasing $a$ and $g$ will enhance the averaged \Ha intensity at all latitudes. Compared with the $\beta$ \hii region model, the $I_{\rm H\alpha}^{\rm tot}$ obtained with the uniform \hii region model shows about $15\%$ lower \Ha intensities at lower latitudes ($|b| \lesssim \SI{20}{\degree}$), but shows about $15\%$ higher \Ha intensities at middle and higher latitudes ($|b| \gtrsim \SI{20}{\degree}$).  For each case, the $1\sigma$ uncertainty in our simulation is primarily caused by the method of setting random distances for \hii regions (see Section \ref{chap:data}) and the process of random scattering, which is typically less than $10\%$ (see Table \ref{Tab2:halpha-sff}).

We further present the scattered \Ha intensity ($I_{\rm H\alpha}^{\rm sca}$) including the scattered emission from the \hii regions ($I_{\rm H\alpha}^{\rm {sca-H{\sc II}}}$) and that from the WIM ($I_{\rm H\alpha}^{\rm {sca-WIM}}$), which is realized by labeling the \Ha photon according to its behavior (i.e., scattered route) in the simulation. As presented in Figure \ref{fig:halpha-plot}, at middle and higher latitudes ($|b| \gtrsim \SI{15}{\degree}$), the $I_{\rm H\alpha}^{\rm {sca-H{\sc II}}}$ increases with the scale radius $R_{\rm s}$, which receives the highest value when the uniform \hii region model is applied. It is found that the scattering percentage is in the range of $15\%$--$50\%$, which is very consistent with the previous observation results, depending on the \hii region model and the scattering parameters of $a$ and $g$. Note that the electron-scattered emission attributes less than $3\%$ of the total scattered \Ha intensity since the cross section of free electrons is three orders of magnitude smaller than that of dust. Therefore, the contribution of scattering caused by the free electrons will no longer be discussed separately.

In addition, we have attempted to employ the cosecant law $I_{\rm H\alpha}^{\rm tot} = A_0 + A_1 / \rm{sin}(|b|)$ to fit the latitudinal distribution of the simulated \Ha intensity, where $A_0$ is the offset, $A_1$ is the amplitude, and $b$ is the Galactic latitude \citep{Dickinson03}. In Figure \ref{fig:halpha-fitting}, we present the best-fitting cosecant profiles of the latitudinal cuts of simulated \Ha intensities (averaged over $\SI{-15}{\degree} \leqslant l \leqslant \SI{15}{\degree}$) for $12$ cases. We find that the cosecant profiles agree well with the simulated total \Ha intensities at lower and middle latitudes ($|b| \lesssim \SI{70}{\degree}$), but they are higher than the $I_{\rm H\alpha}^{\rm {\rm tot}}$ at higher latitudes ($|b| \gtrsim \SI{70}{\degree}$), which are consistent with the simulation result of Wood99. We present the best-fitting cosecant ($A_0$, $A_1$) parameters for $12$ cases in Table \ref{Tab2:halpha-sff}.

\subsubsection{Intrinsic \Ha Intensities}
\label{chap:intrinsic-halpha}

\begin{figure*}[]
	\centering
	\includegraphics[width=0.49\textwidth]{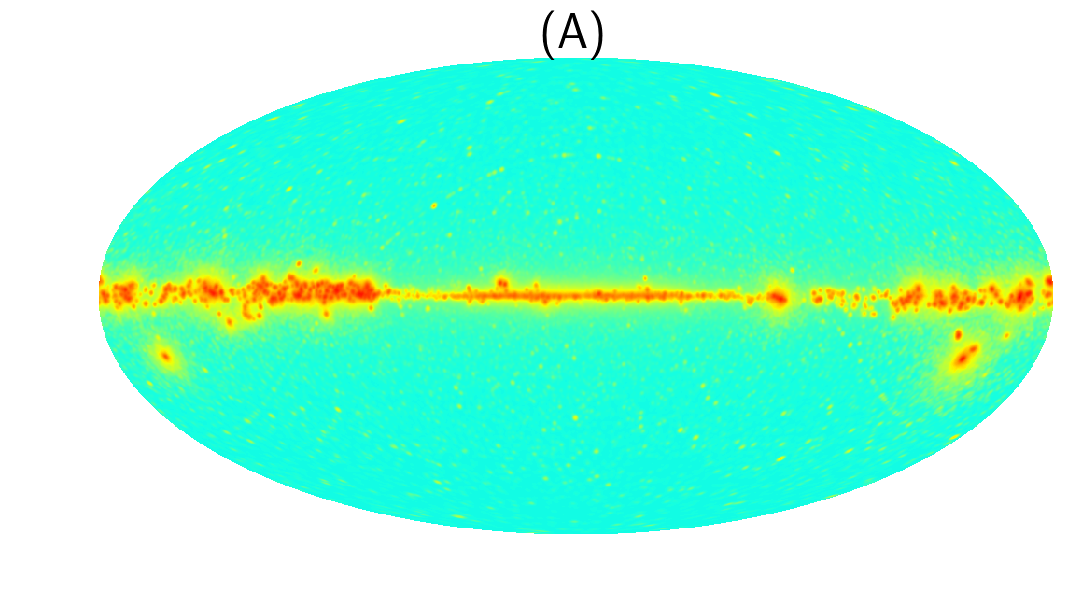}
	\includegraphics[width=0.49\textwidth]{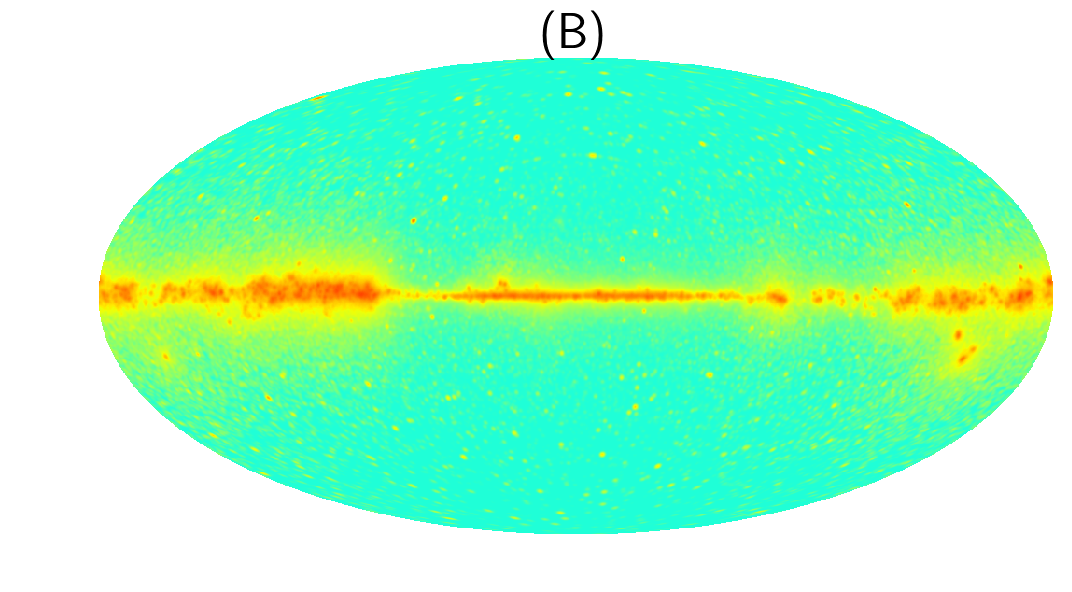}
	\centering
	\includegraphics[width=0.49\textwidth]{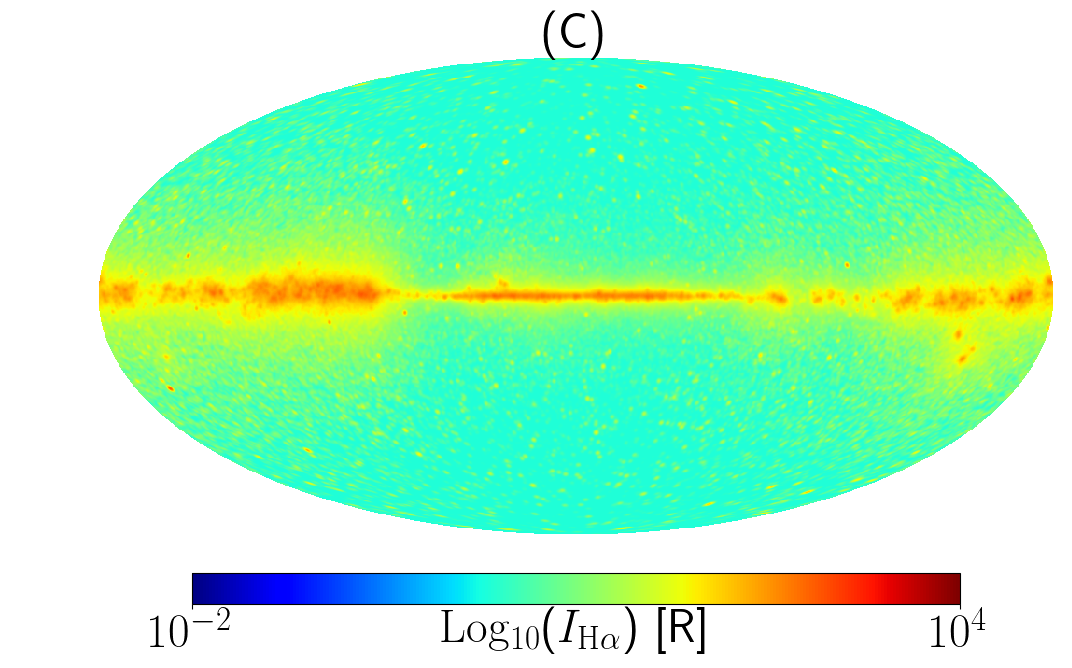}
	\includegraphics[width=0.49\textwidth]{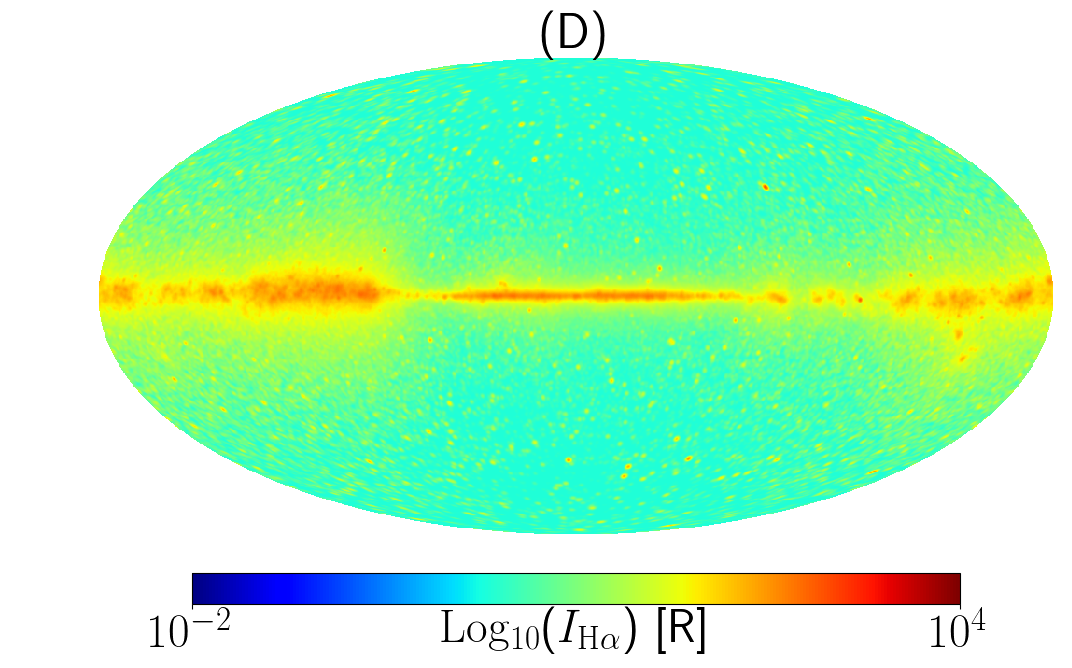}
	\caption{All-sky Mollweide projections ($N_{\rm side} = 1024$) of the intrinsic H$\alpha$ intensity maps simulated with $R_{\rm s} = 0.1 R_{\star}$, $0.5 R_{\star}$, $1.0 R_{\star}$, and uniform \hii models, respectively, in Galactic coordinates with the same figure configuration as Figure \ref{fig:halpha-dust}. All panels show the mean values of $50$ simulation runs and share the same logarithmic scale {\bfseries{in units of $\rm R$.}}}
	\label{fig:halpha-intrinsic-plot}
\end{figure*}

Meanwhile, the intrinsic \Ha intensities are realized by removing the dust and free electrons from the simulation cube. The four cases of intrinsic \Ha intensities ($I_{\rm H\alpha}^{\rm int}$) labeled as (A), (B), (C), and (D) are simulated, which are only relevant to the \hii region models (corresponding to $R_{\rm s} = 0.1 R_{\star}$, $0.5 R_{\star}$, $1.0 R_{\star}$, and uniform \hii region models, respectively.). We present each $I_{\rm H\alpha}^{\rm int}$ map in Figure \ref{fig:halpha-intrinsic-plot}, which is the mean value of $50$ simulation runs. For four intrinsic cases, the $1\sigma$ uncertainties are mainly dominated by the setting random distances for \hii regions (see Section \ref{chap:data}), with a typical value of $\num{\sim 5}\%$ (see the uncertainties listed in Table \ref{Tab2:halpha-sff}).

\begin{figure*}[htbp]
	\centering	
	\includegraphics[width=1.0\textwidth]{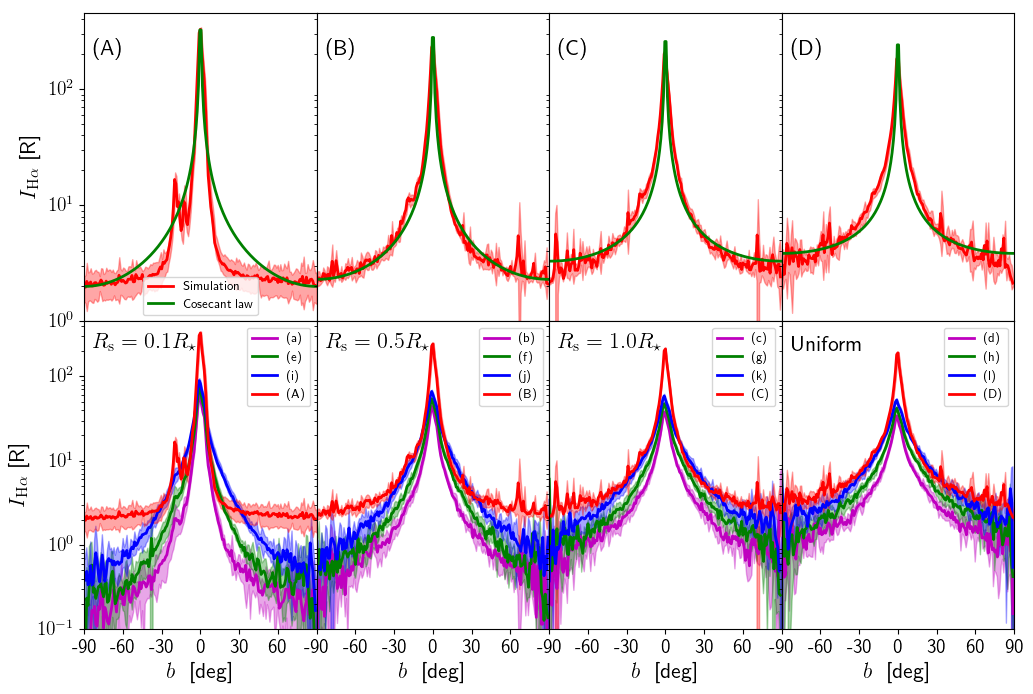}
	\caption{Top panels: latitudinal distributions of the mean (red solid lines) and the $1\sigma$ uncertainties (red shaded regions) of intrinsic \Ha intensities estimated from the $50$ simulation runs, as well as their corresponding cosecant fitting profiles (green solid lines). Bottom panels: comparisons between the simulated total \Ha intensities (magenta, green, and blue solid lines) and the intrinsic \Ha intensities (red solid lines).}
	\label{fig:halpha-intrinsic-fitting}
\end{figure*}

The latitudinal distributions of the means (red solid lines) and $1\sigma$ uncertainties (red shaded regions) of the intrinsic \Ha intensities are shown in Figure \ref{fig:halpha-intrinsic-fitting} (top panels), along with the best-fitting cosecant profiles (green solid lines). We find that the averaged intrinsic \Ha intensities of cases (B), (C), and (D) are about \numlist{1.5; 1.7; 2.0} times more luminous than the $I_{\rm H\alpha}^{\rm int}$ of case (A). It is also found that the $I_{\rm H\alpha}^{\rm int}$ of cases (B), (C), and (D) are consistent with the cosecant law, but the $I_{\rm H\alpha}^{\rm int}$ of case (A) is lower than the cosecant model at middle latitudes ($\SI{10}{\degree} \lesssim |b| \lesssim  \SI{50}{\degree}$), since more \Ha photons are concentrated on the Galactic plane in case (A). The best-fitting cosecant parameters and the averaged intrinsic \Ha intensities are also given in Table \ref{Tab2:halpha-sff}. We further compare the $I_{\rm H\alpha}^{\rm int}$ with the $I_{\rm H\alpha}^{\rm tot}$ and present the results in the bottom panels of Figure \ref{fig:halpha-intrinsic-fitting}. It is found that $I_{\rm H\alpha}^{\rm int}$ is more luminous than $I_{\rm H\alpha}^{\rm tot}$ by about $6.3$, $3.2$, $2.0$, and $1.6$ times when three $\beta$ \hii region models with $R_{\rm s} = 0.1 R_{\star}$, $0.5 R_{\star}$, $1.0 R_{\star}$, and a uniform model of \hii regions are adopted, respectively.

\begin{table*}
	\caption{The Averaged \Ha Intensities with the Cosecant Fitting Parameters of Offset ($A_0$) and Amplitude ($A_1$), and the Corresponding Averaged Galactic Free--Free Brightness Temperatures $T_{\rm b}$ at \numlist{120; 150; 190} \si{\MHz}.}\label{Tab2:halpha-sff}
	\begin{center}
		\setlength{\tabcolsep}{1mm}{
			\begin{tabular}{lccccccccccccccccccccc}
				\hline
				\hline
				Label & ~~~$I_{\rm H\alpha}$ ~~~&~~~$A_0$~~~ &~~~ $A_1$~~~ & ~~$T_{\rm b}$(\SI{120}{\MHz})~~&~~$T_{\rm b}$(\SI{150}{\MHz})~~&~~$T_{\rm b}$(\SI{190}{\MHz}) \\
				&$\left[\rm{R} \right]$ & && $\left[\rm {K}\right]$ &$\left[\rm {K}\right]$&$\left[\rm {K}\right]$\\
				\hline
				(a)& \num{4.43\pm0.30} & \num{-0.70} & \num{0.80} & \num{2.67\pm0.18} & \num{1.67\pm0.11} & \num{1.03\pm0.07}\\
				(b)& \num{4.96\pm0.35} & \num{-0.60} & \num{0.75} & \num{3.00\pm0.21} & \num{1.88\pm0.13} & \num{1.15\pm0.08}\\
				(c)& \num{5.43\pm0.39} & \num{-0.35} & \num{0.70} & \num{3.28\pm0.24} & \num{2.06\pm0.15} & \num{1.26\pm0.09}\\
				(d)& \num{5.78\pm0.40} & \num{0.15}  & \num{0.55} & \num{3.50\pm0.24} & \num{2.20\pm0.15} & \num{1.34\pm0.09}\\
				(e)& \num{6.48\pm0.46} & \num{-0.90} & \num{1.05} & \num{3.91\pm0.28} & \num{2.46\pm0.17} & \num{1.50\pm0.11}\\
				(f)& \num{7.06\pm0.51} & \num{-0.65} & \num{0.95} & \num{4.26\pm0.31} & \num{2.68\pm0.19} & \num{1.64\pm0.12}\\
				(g)& \num{7.65\pm0.55} & \num{-0.25} & \num{0.80} & \num{4.63\pm0.33} & \num{2.91\pm0.21} & \num{1.78\pm0.13}\\
				(h)& \num{8.13\pm0.58} & \num{0.75}  & \num{0.70} & \num{4.92\pm0.35} & \num{3.09\pm0.22} & \num{1.89\pm0.13}\\
				(i)& \num{9.62\pm0.72} & \num{-0.75} & \num{1.15} & \num{5.80\pm0.43} & \num{3.65\pm0.27} & \num{2.23\pm0.17}\\
				(j)& \num{10.13\pm0.74}& \num{-0.65} & \num{1.05} & \num{6.13\pm0.45} & \num{3.86\pm0.28} & \num{2.36\pm0.17}\\
				(k)& \num{10.81\pm0.82}& \num{0.15}  & \num{0.95} & \num{6.54\pm0.50} & \num{4.11\pm0.31} & \num{2.52\pm0.19} \\
				(l)& \num{11.24\pm0.74}& \num{1.05}  & \num{0.85} & \num{6.80\pm0.45} & \num{4.28\pm0.28} & \num{2.62\pm0.17}\\
				\hline
				(A)& \num{17.00\pm0.61} & \num{-0.85} & \num{2.85} & \num{10.07\pm0.36} & \num{6.37\pm0.23} & \num{3.91\pm0.14}\\
				(B)& \num{17.67\pm0.80} & \num{-0.15} & \num{2.45} & \num{10.58\pm0.48} & \num{6.67\pm0.30} & \num{4.08\pm0.18}\\
				(C)& \num{17.76\pm0.75} & \num{1.05}  & \num{2.25} & \num{10.66\pm0.45} & \num{6.72\pm0.28} & \num{4.11\pm0.17}\\
				(D)& \num{18.04\pm0.91} & \num{1.75}  & \num{2.10} & \num{10.84\pm0.55} & \num{6.83\pm0.34} & \num{4.18\pm0.21}\\		
				\hline
				\hline
		\end{tabular}}
	\end{center}
\end{table*}

\subsection{Galactic Free--Free Emission}
\label{chap:free--free}

\begin{figure*}[!hbt]
	\centering
	\includegraphics[width=0.49\textwidth]{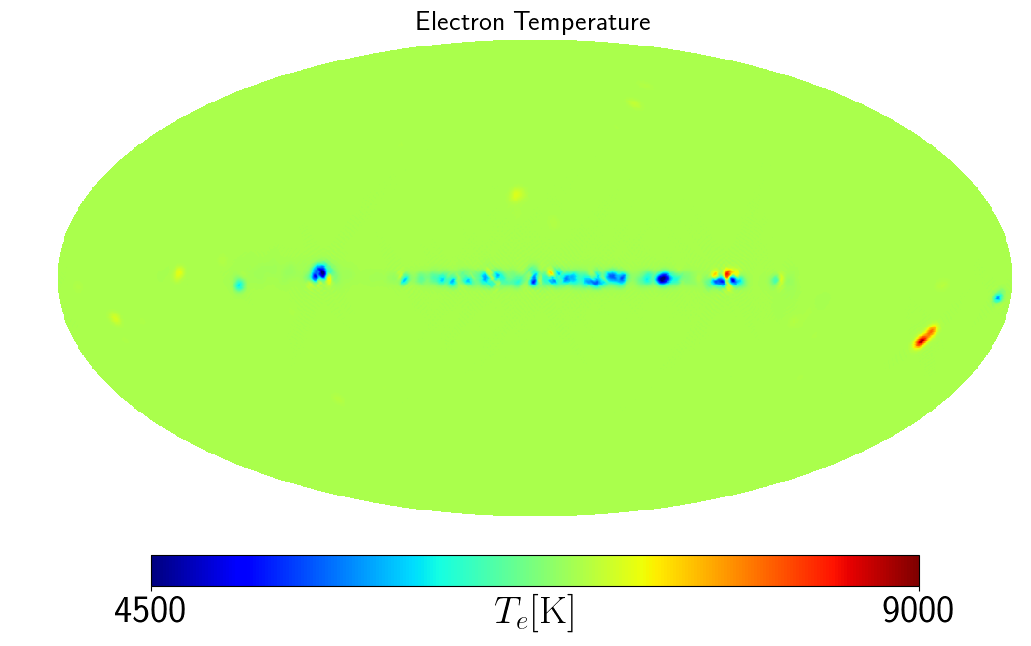}
	\includegraphics[width=0.49\textwidth]{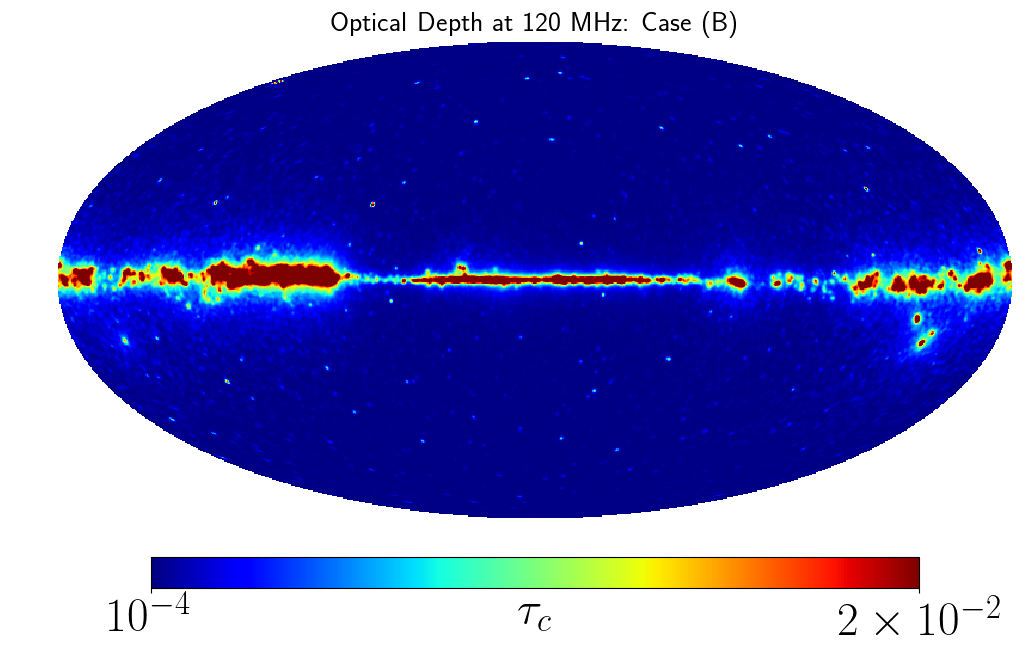}
	\caption{All-sky Mollweide projections ($N_{\rm side} = 1024$) of the electron temperature map (left panel; \citealt{Planck-X16}) reproduced with permission © ESO and an example optical depth map at \SI{120}{\MHz} (right panel) in Galactic coordinates with the same figure configuration as Figure \ref{fig:halpha-dust}. The color bars are in linear scales.} 
	\label{fig:te-tau}
\end{figure*}

We derive the Galactic free--free emission from the above simulated $I_{\rm H\alpha}^{\rm {\rm tot}}$ and $I_{\rm H\alpha}^{\rm int}$ according to the equations given in Section \ref{chap:ff-ha}. To obtain the Galactic free--free emission map, we employ an all-sky electron temperature map proposed by \citet{Planck-X16}, which is presented in the left panel of Figure \ref{fig:te-tau} (reproduced with permission © ESO), and then we can derive the Galactic free--free brightness temperature map at any frequency (\SI{100}{\MHz}--\SI{100}{\GHz}). Meanwhile, an example optical depth map at \SI{120}{\MHz} is shown in the right panel of Figure \ref{fig:te-tau}. For each case, we present the averaged brightness temperatures of Galactic free--free emissions at \numlist{120; 150; 190} \si{\MHz} in Table \ref{Tab2:halpha-sff}.

\begin{figure*}[htbp]
	\centering
	\includegraphics[width=0.325\textwidth]{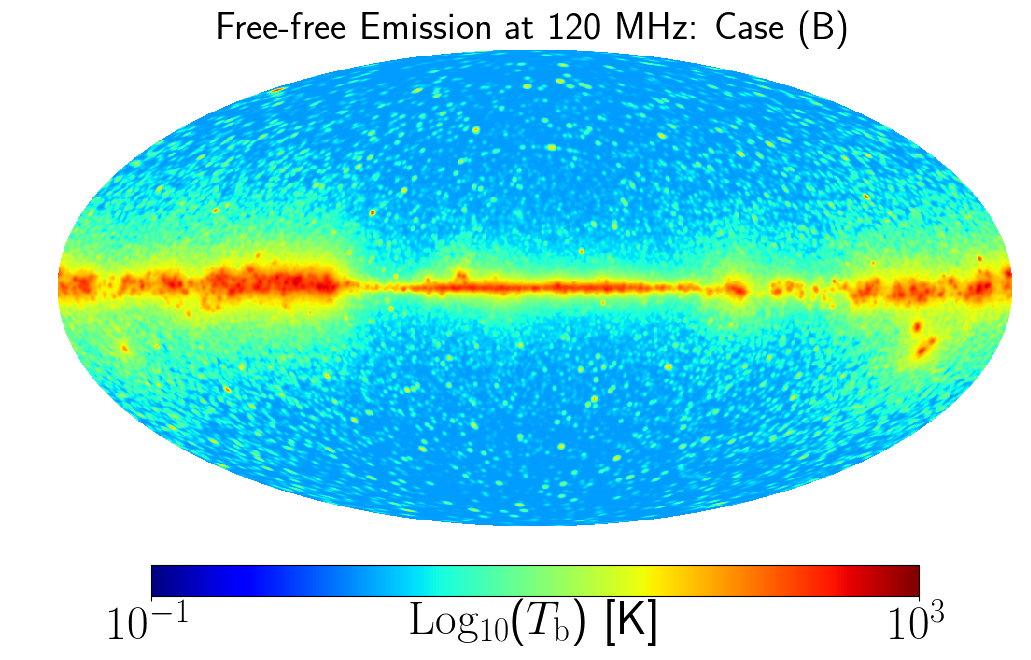}
	\includegraphics[width=0.325\textwidth]{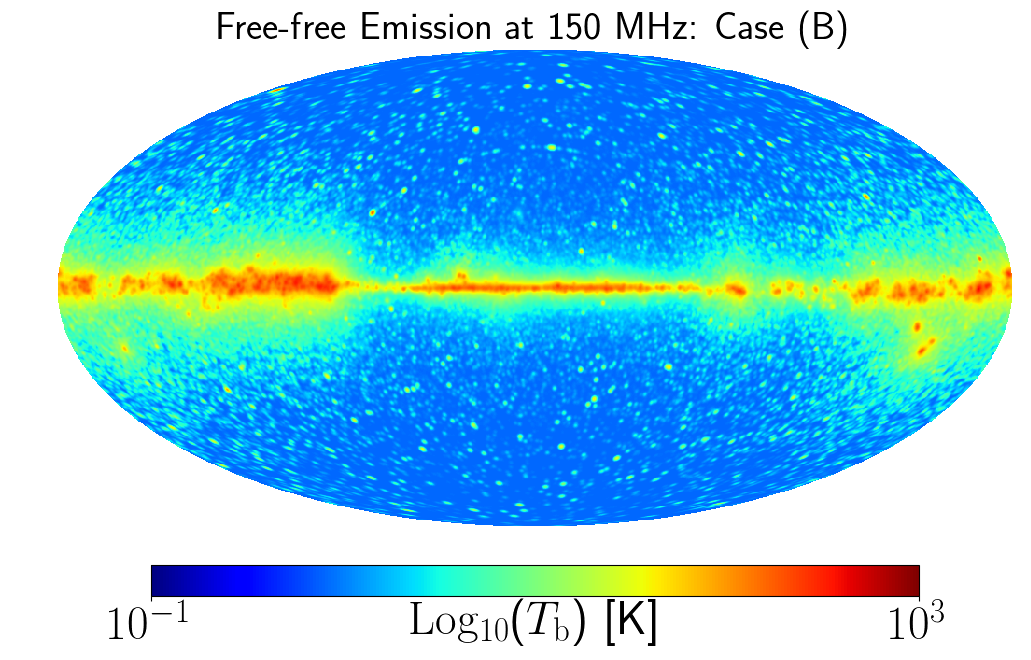}
	\includegraphics[width=0.325\textwidth]{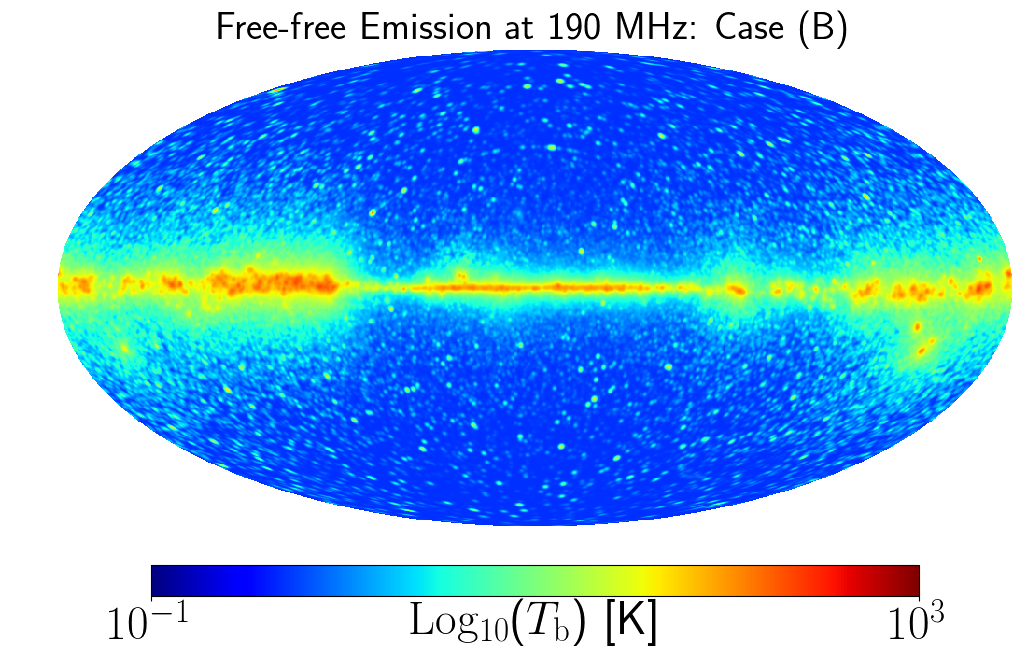}
	\caption{All-sky Mollweide projections ($N_{\rm side} = 1024$) of the Galactic free--free brightness temperatures at \numlist{120; 150; 190} \si{\MHz} in Galactic coordinates with the same figure configuration as Figure \ref{fig:halpha-dust}. All panels show the mean values of $50$ simulation runs and share the same logarithmic scale {\bfseries{in units of $\rm K$.}}} 
	\label{fig:sff}
\end{figure*}

\begin{figure}[htbp]
	\centering
	\includegraphics[width=0.49\textwidth]{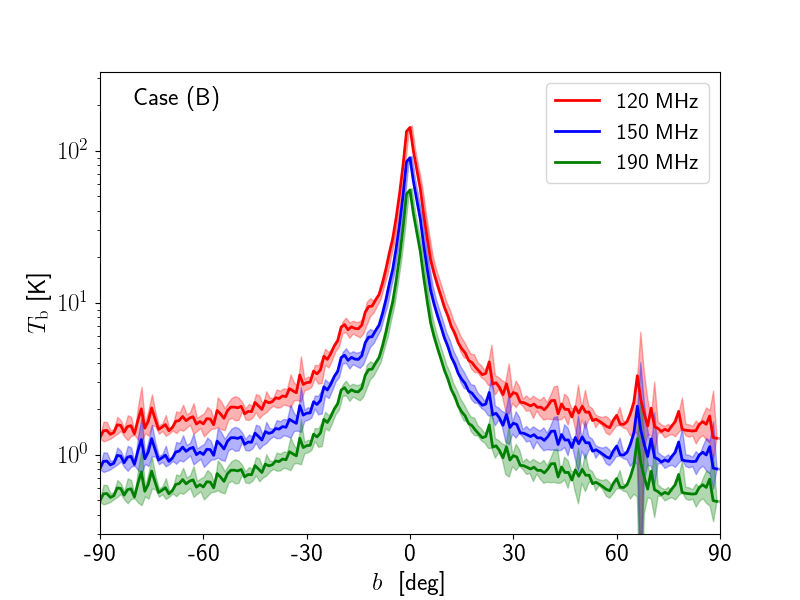}
	\includegraphics[width=0.49\textwidth]{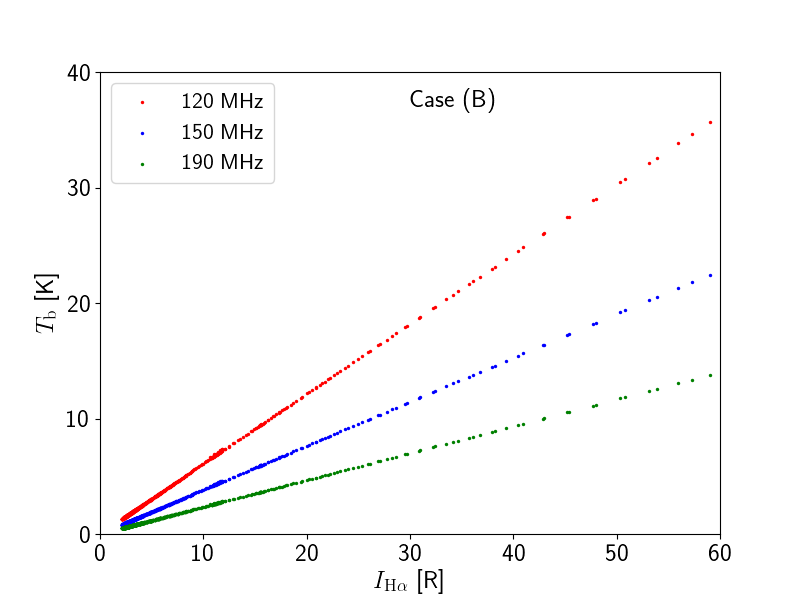}
	\caption{Top panel: latitudinal distributions of Galactic free--free emissions at \numlist{120; 150; 190} \si{\MHz}. The solid lines and shaded regions show the mean values and the corresponding $1\sigma$ uncertainties, respectively. Bottom panel: relations between the \Ha intensities and the Galactic free--free brightness temperatures.}
	\label{fig:sff-lat}
\end{figure}

By comparing the $12$ cases of $I_{\rm H\alpha}^{\rm tot}$ simulated in clumpy dust with the observed \Ha intensity of F03 ($I_{\rm H\alpha}^{\rm F03}$), we recommend the model parameters of case (f), i.e., $a = \num{0.67}$, $g = \num{0.50}$, and $R_{\rm s} = 0.5 R_{\star}$ (for more detailed comparisons, see Section \ref{chap:comp-diss}). Therefore, we derive the Galactic free--free emission from the intrinsic \Ha emission of case (B) to carry out our subsequent calculations. We present the example Galactic free--free emission maps at \numlist{120; 150; 190} \si{\MHz} in Figure \ref{fig:sff}. The latitudinal distributions of the mean and the corresponding $1\sigma$ uncertainties of the Galactic free--free emissions at \numlist{120; 150; 190} \si{\MHz} are presented in Figure \ref{fig:sff-lat} (top panel). The conversions between the Galactic free--free emissions and the corresponding \Ha intensities are illustrated in Figure \ref{fig:sff-lat} (bottom panel), which are \numlist{0.61; 0.38; 0.23} $\rm [K / R]$ at \numlist{120; 150; 190} \si{\MHz}, respectively.

\subsection{Contamination of Galactic Free--Free Emission}
\label{chap:Gff2EoR}

The 1D and 2D power spectra are calculated to estimate the contamination of Galactic free--free emission on the EoR signal. The EoR signals observed at different frequencies are expected to be a 3D image cube, where the two angular dimensions describe the transverse distances across the sky and the one frequency dimension depicts the line-of-sight distance. For each foreground component cube, its two angular dimensions describe the same sky coverage as the EoR signal, but its one frequency dimension depicts the emission distribution in the frequency space (i.e., spectrum), which is different from the EoR signal. The 3D power spectrum $P(k_x, k_y, k_z)$ of the EoR signal should be spherical symmetry within a limited redshift range (e.g., $\Delta z \sim 0.5$, corresponding to a frequency bandwidth of $\sim$ \SI{8}{\MHz} at \SI{150}{\MHz}), during which the evolution of the universe can be ignored and the \hi can be regarded as isotropic. The spherically averaged 1D $k$-space power spectrum $P(k)$ can be calculated by averaging the $P(k_x, k_y, k_z)$ to achieve a relatively higher signal-to-noise ratio. As adopted in both the theoretical studies (e.g., \citealt{Morales04,Datta10}) and the low-frequency experiments (e.g., \citealt{Li19}), the dimensionless variant of the 1D power spectrum $\Delta^{2}(k)$ = $P(k)k^{3}/(2\pi^{2})$ is more commonly employed. The Blackman--Nuttall window function is applied to the frequency dimension before calculating the 3D power spectra to suppress the significant sidelobes in the Fourier transform \citep{Trott15,Chapman16,Li19}.

\begin{figure*}[htbp]             
	\centering
	\includegraphics[width=1.0\textwidth]{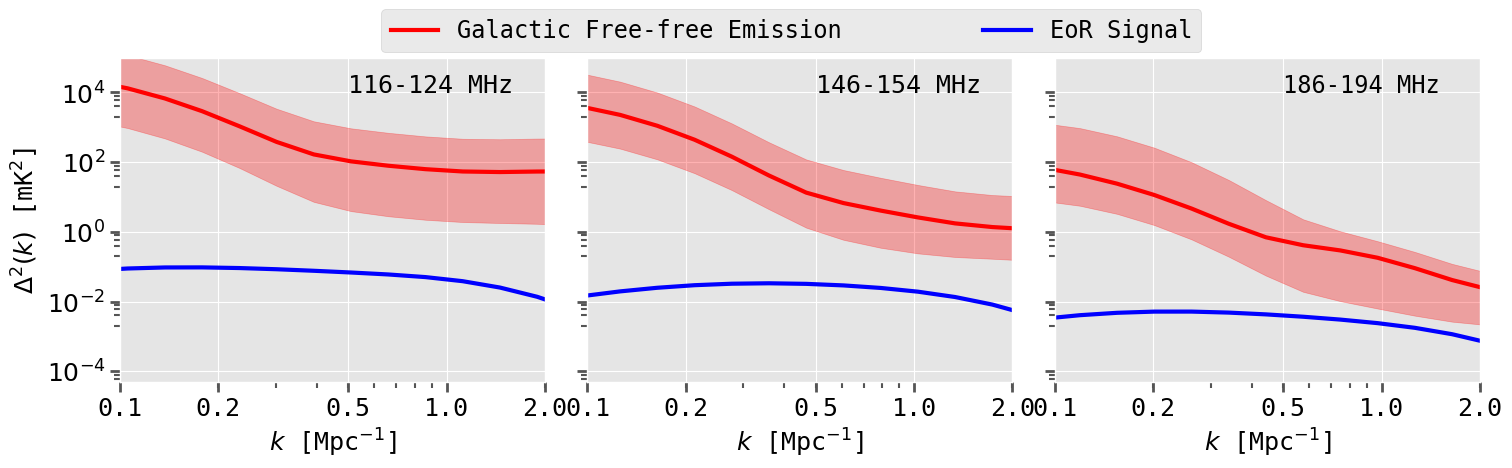}	
	\caption{The 1D power spectra $\Delta^{2}(k)$ of Galactic free--free emission (red solid line) and the EoR signal (blue solid line) in the \numrange{116}{124}, \numrange{146}{154}, and \numrange{186}{194} MHz frequency bands. The red solid lines and red shaded regions show the mean values and the corresponding $1\sigma$ uncertainties of the power spectra of Galactic free--free emission estimated from the $50$ simulation runs.}
	\label{fig:ps1d-comp}
\end{figure*}

We calculate the 1D power spectra $\Delta^{2}(k)$ from the SKA \enquote{observed} image cubes of the Galactic free--free emission and the EoR signal. The comparisons of the power spectra $\Delta^{2}(k)$ between the Galactic free--free emission and the EoR signal are presented in Figure \ref{fig:ps1d-comp}. It is obvious that the contamination caused by Galactic free--free emission on the EoR signal is a function of position in the $k$-space. On large scales ($k \lesssim \SI{0.5}{\per\Mpc}$) the Galactic free--free emission has a greater impact on the EoR signal, while on small scales ($k \gtrsim \SI{0.5}{\per\Mpc}$) it causes relatively less contamination on the EoR signal. We find that, given the $1\sigma$ uncertainties, the Galactic free--free emissions are more luminous than the EoR signals by about \numrange{e5.4}{e2.1}, \numrange{e5.0}{e1.7}, and \numrange{e4.3}{e1.1} times on scales of $\SI{0.1}{\per\Mpc} < k < \SI{2}{\per\Mpc}$ in the \numrange{116}{124}, \numrange{146}{154}, and \SIrange{186}{194}{\MHz} frequency bands, respectively.

\begin{figure*}[htbp]               
	\centering
	\includegraphics[width=1.0\textwidth]{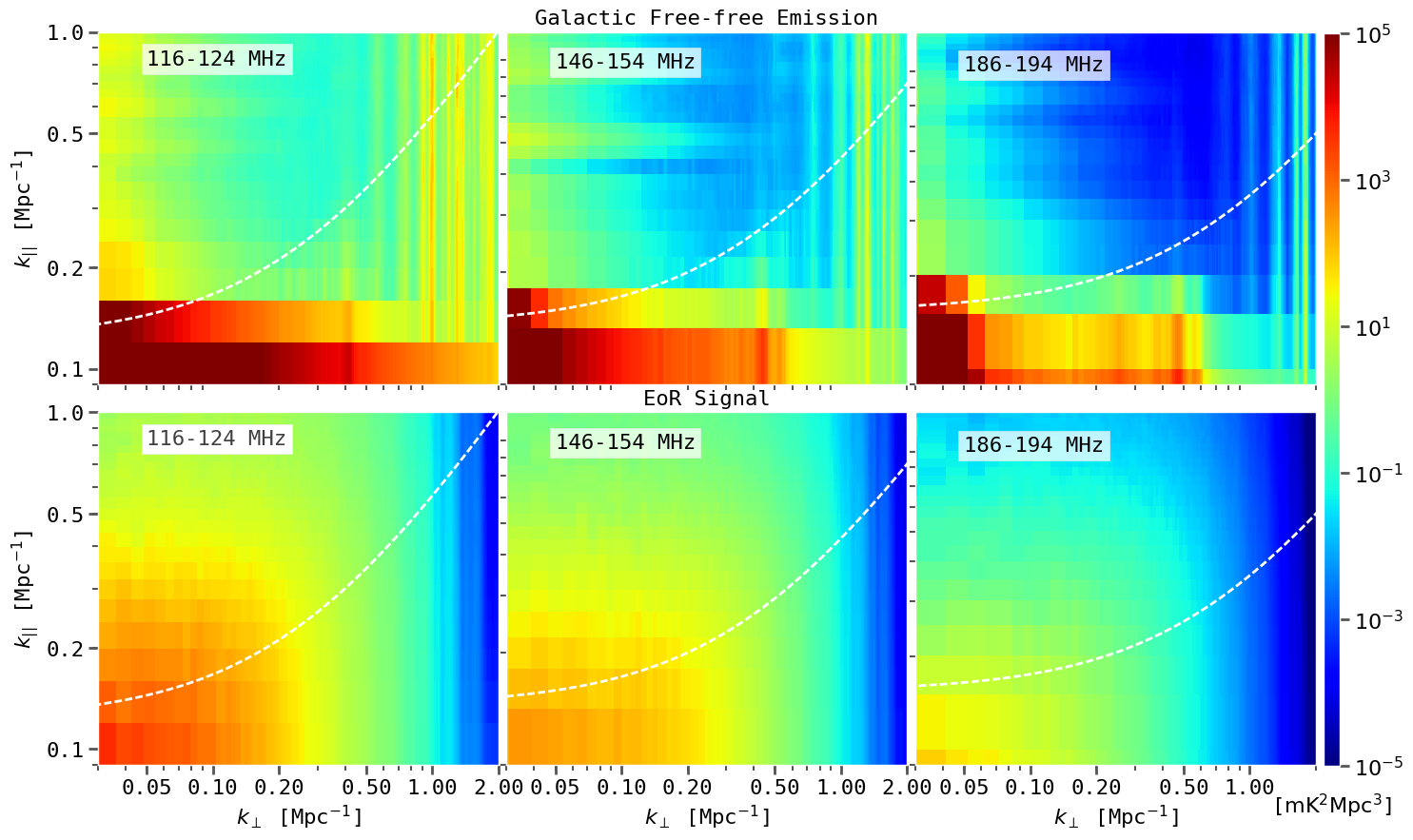}	
	\caption{The 2D power spectra $P(k_{\perp}, k_{||})$ of Galactic free--free emission (mean of $50$ simulation runs; top panels) and the EoR signal (bottom panels) in the \numrange{116}{124}, \numrange{146}{154}, and \SIrange{186}{194}{\MHz} frequency bands. The white dashed lines mark the boundary between the foreground wedge (at the bottom right) and the EoR window (at the top left). All panels share the same logarithmic scale in units of $\rm {mK^{2}Mpc^{3}}$.}
	\label{fig:ps2d}
\end{figure*}

The 2D power spectrum $P(k_{\perp}, k_{||})$ can be obtained by averaging the 3D power spectrum $P (k_x, k_y, k_z)$ over the corresponding angular annuli, the radius of which is $k_{\perp} \equiv \sqrt{k_{x}^{2} + k_{y}^{2}}$, for each line-of-sight plane $k_{||} \equiv k_z$. It is found that in the ($k_{\perp}, k_{||}$) plane the spectral-smooth Galactic free--free emission dominates the low-$k_{||}$ region, but some purely angular ($k_{\perp}$) modes of the foreground signal can be thrown into the line-of-sight ($k_{||}$) dimension (called mode mixing), due to the complicated instrumental and observational effects (e.g., chromatic primary beams, calibration errors). Consequently, an expanded wedge-like contamination region appears at the bottom right in the ($k_{\perp}, k_{||}$) plane, which is known as the foreground wedge \citep{Datta10,Morales12,Liu14}. The top left corner in the ($k_{\perp}, k_{||}$) plane, on the other hand, is almost free from the foreground contamination, namely, the EoR window, whose description is proposed by \citet{Thyagarajan13}

\begin{equation}
\label{equ:EoR}
k_{||} \geq \frac{H(z)D_{\rm M}(z)}{(1+z)c} [k_{\perp}~{\rm {sin} }\Theta + \frac{2\pi w f_{21}}{(1 + z) D_{\rm M}(z) B}]
\end{equation}

\noindent where $H(z)$ is the Hubble parameter at redshift $z$, $D_{\rm M}(z)$ is the transverse comoving distance, $B = \SI{8}{\MHz}$ is the frequency bandwidth of the image cube, $w$ ($\propto B$) is the number of characteristic convolution widths for the spillover region caused by the variations in instrumental frequency response, $\Theta$ is the angular distance of the foreground sources from the field center, and $f_{21} = \SI{1420.4}{\MHz}$ is the rest frequency of the 21 cm emission line.

We calculate the 2D power spectra $P(k_{\perp}, k_{||})$ of the Galactic free--free emission and the EoR signal in the \numrange{116}{124}, \numrange{146}{154}, and \SIrange{186}{194}{\MHz} frequency bands and present the results in Figure \ref{fig:ps2d}. We find that the spectral-smooth Galactic free--free emission dominates the low-$k_{||}$ ($k_{||}$ $\lesssim$ \SI{0.2}{\per\Mpc}) regions, while the EoR signal distributes its power across all $k_{||}$ modes, illustrating its rapid fluctuations along the line-of-sight dimension. Concerning the angular dimension, the powers of Galactic free--free emission and the EoR signal dominate on scales of $k_{\perp}$ $\lesssim$ \SI{0.2}{\per\Mpc}.

\begin{figure*}[htbp]                
	\centering
	\includegraphics[width=1.0\textwidth]{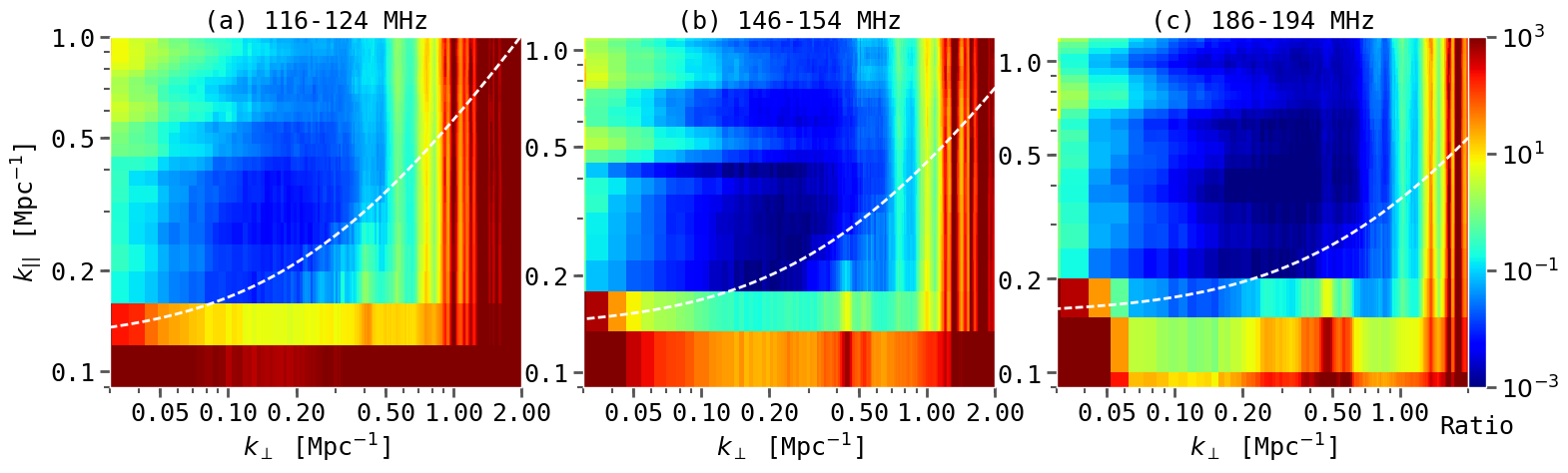}	
	\caption{The 2D power spectra ratios $R(k_{\perp}, k_{||})$ of Galactic free--free emission to the EoR signal in the \numrange{116}{124}, \numrange{146}{154}, and \SIrange{186}{194}{\MHz} frequency bands. The mean 2D power spectrum of $50$ simulation runs for Galactic free--free emission is used. The white dashed lines mark the EoR window boundaries.}
	\label{fig:ps2d-ratio}
\end{figure*}

To better constrain the contamination caused by Galactic free--free emission, we then calculate the 2D power spectrum ratio $R(k_{\perp}, k_{||})$ defined as $R(k_{\perp}, k_{||})$ = $P_{\rm Gff}(k_{\perp}, k_{||})$ / $P_{\rm 21cm}(k_{\perp}, k_{||})$, where $P_{\rm Gff}(k_{\perp}, k_{||})$ and $P_{\rm 21cm}(k_{\perp}, k_{||})$ are the 2D power spectra of the Galactic free--free emission and the EoR signal, respectively. As presented in Figure \ref{fig:ps2d-ratio}, the EoR signal is almost free from the contamination of Galactic free--free emission on scales of $k_{||}$ $\gtrsim$ \SI{0.17}{\per\Mpc} and $k_{\perp}$ $\lesssim$ \SI{0.5}{\per\Mpc}, $k_{||}$ $\gtrsim$ \SI{0.19}{\per\Mpc} and $k_{\perp}$ $\lesssim$ \SI{0.7}{\per\Mpc}, and $k_{||}$ $\gtrsim$ \SI{0.2}{\per\Mpc} and $k_{\perp}$ $\lesssim$ \SI{0.9}{\per\Mpc}, in the \numrange{116}{124}, \numrange{146}{154}, and \SIrange{186}{194}{\MHz} frequency bands, respectively, while outside these regions, the Galactic free--free emission causes significant contamination, because the 2D power spectrum ratio is obviously greater than unity in three frequency bands.

\begin{figure}[htbp]          
	\centering
	\includegraphics[width=0.48\textwidth]{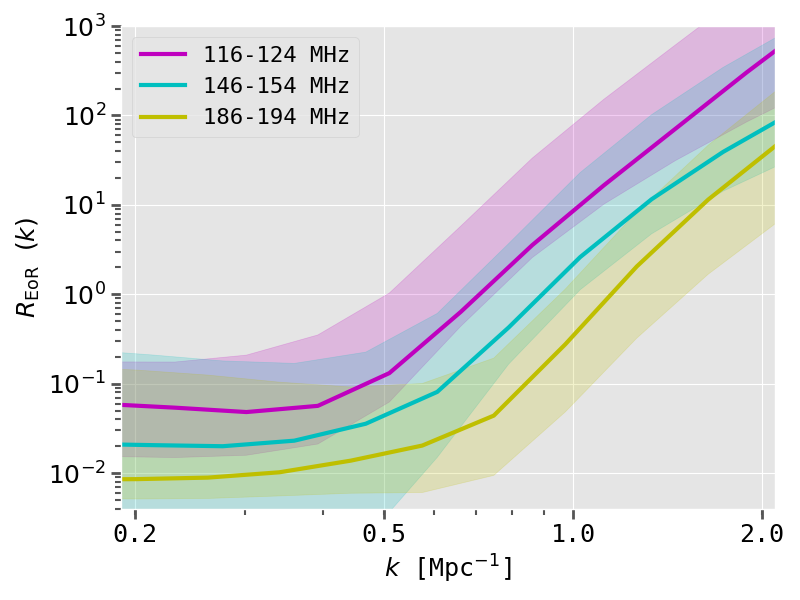}		
	\caption{The 1D power ratios $R_{\rm EoR}(k)$ inside the EoR window of Galactic free--free emission to the EoR signal in the \numrange{116}{124}, \numrange{146}{154}, and \SIrange{186}{194}{\MHz} frequency bands. The solid lines and shaded regions show the mean values and the corresponding $1\sigma$ uncertainties, respectively.} 
	\label{fig:ps1d-eor-ratio}
\end{figure}

To further quantify the contamination imposed by Galactic free--free emission, we define an EoR window (marked by white dashed lines in Figure \ref{fig:ps2d} and Figure \ref{fig:ps2d-ratio}) in the ($k_{\perp}, k_{||}$) plane according to Equation \ref{equ:EoR} with a configuration of $w = 3$ and the SKA1-Low's FOV (i.e., $\Theta = \SI{6}{\degree}$, $\SI{5}{\degree}$, and $\SI{4}{\degree}$ in the \numrange{116}{124}, \numrange{146}{154}, and \SIrange{186}{194}{\MHz} frequency bands, respectively). We then calculate the 1D power spectrum ratio $R_{\rm EoR}(k)$ of Galactic free--free emission to the EoR signal by averaging the modes only inside the EoR window. As shown in Figure \ref{fig:ps1d-eor-ratio}, inside the EoR window, the impact induced by the leaked Galactic free--free emission on the EoR signal can be ignored on large scales ($k$ $\lesssim$ \SI{0.5}{\per\Mpc}), while the leaked Galactic free--free emission causes severe contamination on the EoR detection on small scales ($k$ $\gtrsim$ \SI{0.5}{\per\Mpc}). These results are consistent with the analysis of 2D power spectrum ratios (see Figure \ref{fig:ps2d-ratio}). We find that compared to Figure \ref{fig:ps1d-comp}, the 1D power ratios inside the EoR window $R_{\rm EoR}$($k$) are suppressed by about $3$ orders of magnitude, which illustrates that the EoR window is a powerful tool in detecting the EoR signal. For example, on scales of $k \sim \SI{0.5}{\per\Mpc}$, the $R_{\rm EoR}$($k$) are generally about $12\%$, $5\%$, and $2\%$ in the \numrange{116}{124}, \numrange{146}{154}, and \SIrange{186}{194}{\MHz} frequency bands, respectively. However, even inside the EoR window, the power leaked by Galactic free--free emission can still be significant, as the $R_{\rm EoR}$($k$) can be up to about $110\%$--$8000\%$, $30$\%--$2400\%$, and $10\%$--$250\%$ when considering the $1\sigma$ uncertainties (shaded regions) on scales of $\SI{0.5}{\per\Mpc} \lesssim k \lesssim \SI{1}{\per\Mpc}$ in the three frequency bands, respectively. These analyses further support that the Galactic free--free emission should be carefully removed in the EoR detections, especially toward the lower frequencies (\SI{\sim 116}{\MHz}).

\section{Comparison and Discussion}
\label{chap:comp-diss}

\begin{figure*}[htbp]
	\centering
	\includegraphics[width=1.0\textwidth]{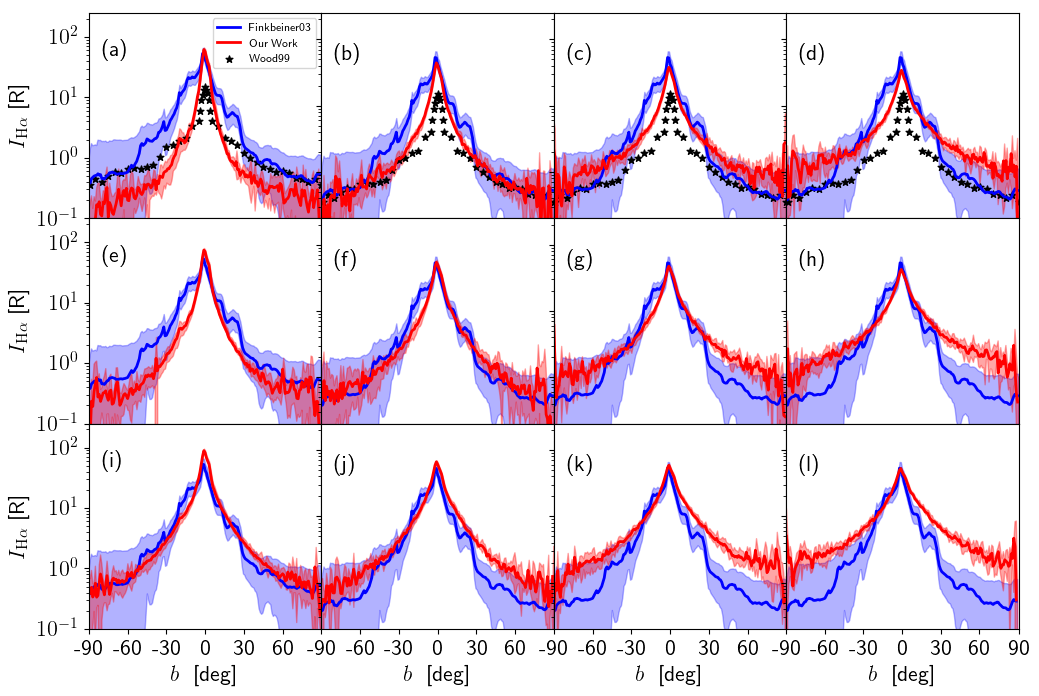}	
	\caption{Comparisons between the simulated total \Ha intensities (red solid lines) and the observed \Ha intensities of F03 (blue solid lines). The mean values of $50$ simulation runs are used for the simulated total \Ha intensities. The red-shaded regions show the corresponding $1\sigma$ uncertainties of the $I_{\rm H\alpha}^{\rm tot}$, and the blue-shaded regions show the errors of $I_{\rm H\alpha}^{\rm F03}$. The black asterisks mark the simulated \Ha intensities of Wood99.}
	\label{fig:halpha-compare}
\end{figure*}

To quantitatively verify our simulation, we compare the simulated $I_{\rm H\alpha}^{\rm tot}$ of cases (a), (b), (c), and (d) with the result of Wood99, given that they share the same scattering parameters ($a = 0.50$, $g = 0.44$). The black asterisks in Figure \ref{fig:halpha-compare} mark the simulated Wood99 \Ha intensity, which is about $25\%$, $35\%$, and $45\%$ lower than $I_{\rm H\alpha}^{\rm tot}$ of cases (b), (c), and (d), respectively. The \Ha intensity of case (a) is about $25\%$ higher than that of Wood99 at lower latitudes ($|b| \lesssim \SI{10}{\degree}$), but it is about $35\%$ lower at middle and higher latitudes ($|b| \gtrsim \SI{10}{\degree}$). The departures between them are due to the different \hii region models and different dust models, as in our simulation the \hii regions are modeled with detailed $\beta$ or uniform structures (see Section \ref{chap:data}) other than just simply \enquote{point sources}. Moreover, the clumpy dust is derived from the observed \hi data rather than a simple axisymmetric model.

\begin{figure*}[htbp]
	\centering
	\includegraphics[width=0.325\textwidth]{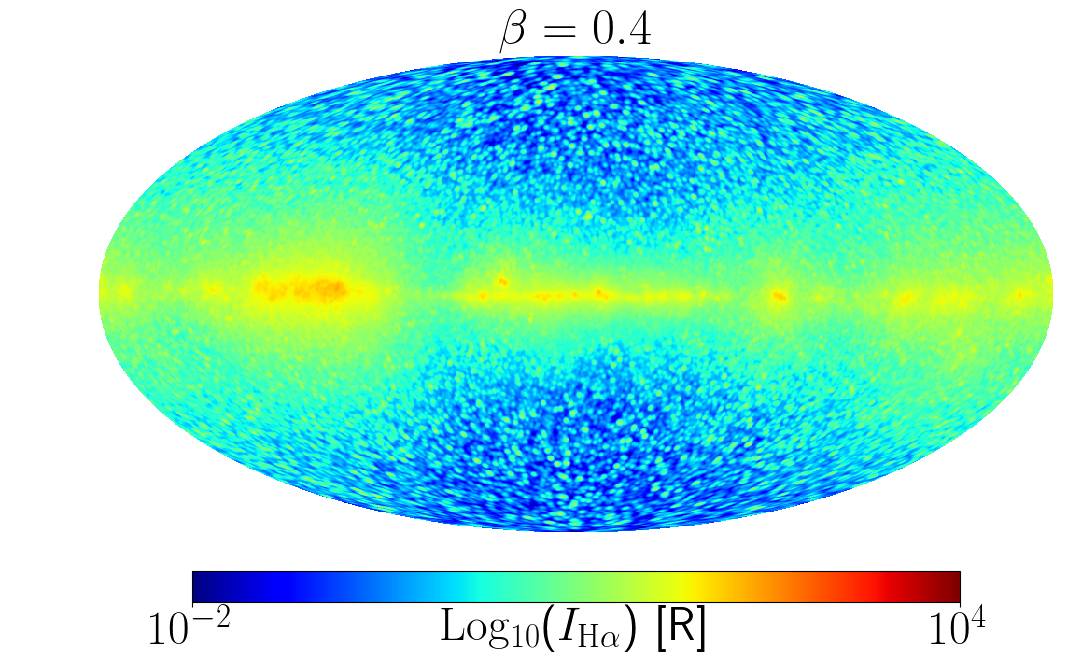}
	\includegraphics[width=0.325\textwidth]{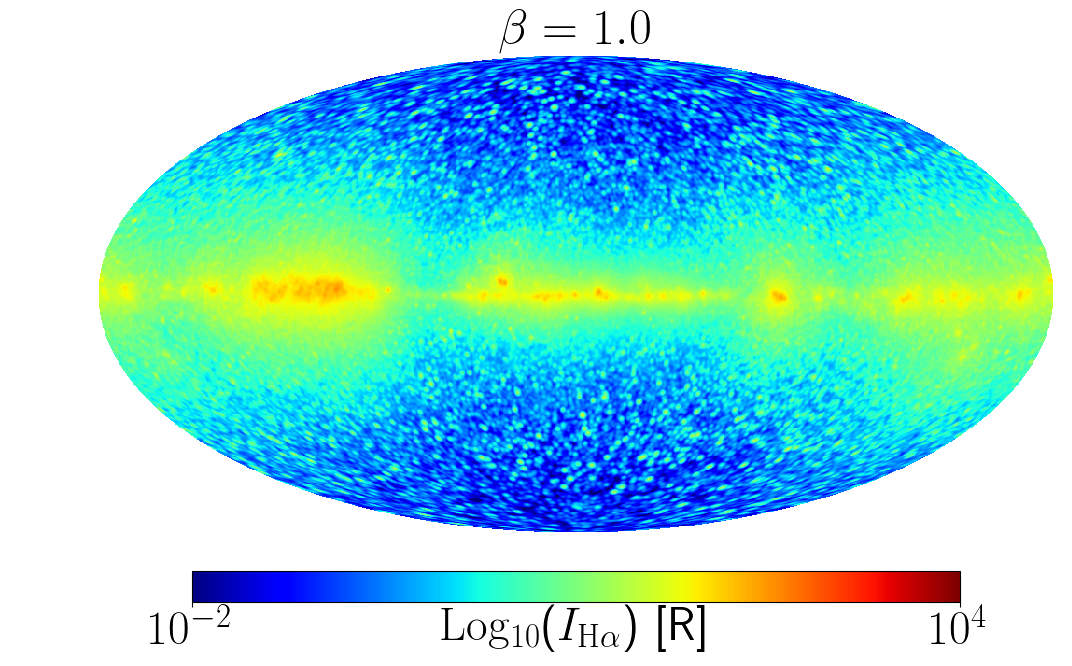}
	\includegraphics[width=0.325\textwidth]{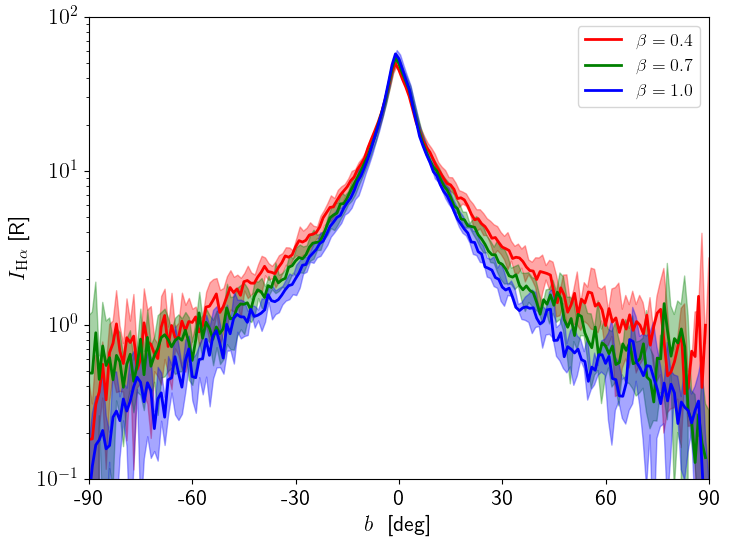}
	\caption{All-sky Mollweide projections ($N_{\rm side} = 1024$) of the simulated total \Ha intensity maps modeled with $\beta = 0.4$ (left panel) and $\beta = 1.0$ (middle panel) in Galactic coordinates with the same figure configuration as Figure \ref{fig:halpha-dust}. The left and middle panels show the mean values of 50 simulation runs and share the same logarithmic scale in units of $\rm R$. Right panel: corresponding comparisons of the simulated \Ha intensities among three simulations obtained with $\beta = 0.4$, $0.7$, and $1.0$. The solid lines and shaded regions show the mean values and the corresponding $1\sigma$ uncertainties estimated from $50$ simulation runs, respectively.} 
	\label{fig:halpha-diss}
\end{figure*}

F03 derived an all-sky observed \Ha intensity map \footnote{\url{https://faun.rc.fas.harvard.edu/dfink/skymaps/}} by jointly studying three \Ha surveys, i.e., Virginia Tech Spectral line Survey (VTSS \footnote{\url{http://www.phys.vt.edu/~halpha}}; \citealt{Dennison98}), SHASSA \footnote{\url{http://amundsen.swarthmore.edu/SHASSA}} (\citealt{Gaustad01}), and WHAM \footnote{\url{http://www.astro.wisc.edu/wham}} (\citealt{Haffner03}). For each case, we further compare the simulated total \Ha intensity map with the observed F03 \Ha intensity map and present the mean and $1\sigma$ uncertainty of $I_{\rm H\alpha}^{\rm tot}$, as well as the value and corresponding error of $I_{\rm H\alpha}^{\rm F03}$ in Figure \ref{fig:halpha-compare}. The error of the $I_{\rm H\alpha}^{\rm F03}$ is caused by the calibration uncertainty in bright regions, readout noise, and foreground Poisson errors in faint regions. As shown in Figure \ref{fig:halpha-compare}, the $I_{\rm H\alpha}^{\rm tot}$ of cases (a), (b), (c), and (d) are about $45\%$, $38\%$, $32\%$, and $28\%$ lower than the $I_{\rm H\alpha}^{\rm F03}$, respectively, and the departures between them are due to the fact that less scattered \Ha intensities are produced for their smaller scattering parameters ($a = 0.50$, $g = 0.44$). The case (e) shows about $\num{\sim 19}\%$ lower \Ha intensity, because more \Ha photons in simulation with $R_{\rm s} = 0.1 R_{\star}$ reside on the Galactic plane, where the absorption effect is severest. The $I_{\rm H\alpha}^{\rm tot}$ of case (f) is in qualitative agreement with the $I_{\rm H\alpha}^{\rm F03}$ at latitudes of $\SI{-90}{\degree} \lesssim b \lesssim \SI{-30}{\degree}$ and $\SI{-10}{\degree} \lesssim b \lesssim \SI{90}{\degree}$, with slightly ($\num{\sim 5}\%$) lower values at latitudes of $\SI{-30}{\degree} \lesssim b \lesssim \SI{-10}{\degree}$. The averaged $I_{\rm H\alpha}^{\rm tot}$ of case (f) is \num{7.06\pm0.51} ${\rm R}$, which is also comparative with the observed value \num{8.04\pm1.29} ${\rm R}$ proposed by F03. Based on the above analyses, we recommend the parameters of model (f) ($\beta = \num{0.7}$, $R_{\rm s} = 0.5 R_{\star}$, $a = \num{0.67}$, and $g = \num{0.50}$). Cases (g) and (h) show $\num{\sim 18}\%$ and $\num{\sim 25}\%$ higher \Ha intensities at middle and higher latitudes ($|b| \gtrsim \SI{25}{\degree}$), as more \Ha photons will scatter to middle and higher latitudes. Cases (i), (j), (k), and (l) show about $20\%$, $26\%$, $34\%$, and $40\%$ higher \Ha intensities, respectively, because more scattered \Ha intensities are obtained owing to the larger scattering parameters ($a = \num{0.77}$, $g = \num{0.55}$).

We also compare simulations with three different $\beta = \num{0.4}$, $\num{0.7}$ (case (f); see Figure \ref{fig:halpha-dust}), and $\num{1.0}$ (take $R_{\rm s} = 0.5 R_{\star}$, $a = \num{0.67}$, and $g = \num{0.50}$, for instance) to test the effect of $\beta$ on the simulated total \Ha intensity, and we present the results in Figure \ref{fig:halpha-diss}. The averaged $I_{\rm H\alpha}^{\rm tot}$ for simulations with $\beta = \num{0.4}$, $\num{0.7}$, and $\num{1.0}$ are \num{7.47\pm0.54}, \num{7.06\pm0.51}, and \num{6.79\pm0.48}, respectively. The comparisons of latitudinal distributions of the $I_{\rm H\alpha}^{\rm tot}$ simulated with $\beta = \num{0.4}$, $\num{0.7}$, and $\num{1.0}$ are shown in the right panel of Figure \ref{fig:halpha-diss}. Compared with case (f) ($\beta = \num{0.7}$), we find that the $I_{\rm H\alpha}^{\rm tot}$ at lower latitudes ($|b| \lesssim \SI{10}{\degree}$) increases slightly with the increase of $\beta$, while it shows a contrary tendency at middle and higher latitudes ($|b| \gtrsim \SI{10}{\degree}$). In conclusion, we argue that the uncertainty of $I_{\rm H\alpha}^{\rm tot}$ caused by $\beta$ (when the $\beta$ changes from $0.7$ to $0.4$ or from $0.7$ to $1.0$) is less than $5\%$.

\section{Summary}
\label{chap:summary}

We have implemented an all-sky Galactic free--free emission map based on the Monte Carlo simulation of the \Ha intensity incorporating the direct and scattered emissions from \hii regions and the WIM. Our simulation recovers the main structures of the Milky Way and reproduces the major characteristics of the observed \Ha intensity - the cosecant profile. We finally recommend a set of model parameters of $\beta = 0.7$, $R_{\rm s} = 0.5 R_{\star}$, $a = \num{0.67}$, and $g = \num{0.50}$ to match the current observation data. Based on the intrinsic \Ha intensity, we derive the Galactic free--free emission and evaluate its contamination on the EoR detection, for which we have incorporated the instrumental effects by utilizing the latest SKA1-Low layout configuration. By carrying out detailed comparisons of the power spectra between Galactic free--free emission and the EoR signal in the \numrange{116}{124}, \numrange{146}{154}, and \SIrange{186}{194}{\MHz} frequency bands, we have shown that the contamination of Galactic free--free emission on the EoR signal is a function of position in the $k$-space, i.e., on large scales ($k$ $\lesssim$ \SI{0.5}{\per\Mpc}) the Galactic free--free emission causes severe contamination, especially toward lower frequencies (\SI{\sim 116}{\MHz}), while on small scales ($k$ $\gtrsim$ \SI{0.5}{\per\Mpc}) it causes relatively less contamination on the EoR detection. Even inside the properly defined EoR window, the power leaked by Galactic free--free emission can still cause nonnegligible contamination on the EoR signal. Overall, we recommend that the Galactic free--free emission, as a severe contaminating source, needs serious treatment in the forthcoming deep EoR experiments.

\section*{Acknowledgments}
We are very grateful to the reviewer for the constructive comments that greatly helped improve the manuscript.
We thank Jayant Murthy and M. S. Akshaya for their suggestions and the code to obtain the 3D distribution of the \hi density.
All simulations are performed on the high-performance cluster at the Department of Astronomy, Shanghai Jiao Tong University.
This work is supported by the Ministry of Science and Technology of China (grant No. 2018YFA0404601)
and the National Natural Science Foundation of China (grant Nos. 11621303, 11835009, and 11973033).


\end{document}